\documentclass[pra,print,superscriptaddress,twocolumn]{revtex4}
\usepackage{amssymb}
\usepackage{amsmath}
\usepackage{epsfig}
\usepackage{color}
\usepackage{graphics, graphicx}
\usepackage{bbold}
\usepackage{psfrag}
\usepackage{mathcomp}
\usepackage{subfigure}
\usepackage{verbatim}
\usepackage{color}
\usepackage[colorlinks,citecolor=blue]{hyperref}

\begin{document}

\title{Collective dynamics of the unbalanced three-level Dicke model}
\author{Jingtao Fan}
\thanks{fanjt@sxu.edu.cn}
\affiliation{State Key Laboratory of Quantum Optics and Quantum Optics Devices, Institute
of Laser Spectroscopy, Shanxi University, Taiyuan 030006, China}
\affiliation{Collaborative Innovation Center of Extreme Optics, Shanxi University,
Taiyuan 030006, China}
\author{Suotang Jia}
\affiliation{State Key Laboratory of Quantum Optics and Quantum Optics Devices, Institute
of Laser Spectroscopy, Shanxi University, Taiyuan 030006, China}
\affiliation{Collaborative Innovation Center of Extreme Optics, Shanxi University,
Taiyuan 030006, China}

\begin{abstract}
We study a three-level Dicke model in V-configuration under both closed and
open conditions. With independently tunable co- and counter-rotating
coupling strength of the interaction Hamiltonian, this model is a
generalization of the standard Dicke model that features multiple distinct
parameter regimes. Based on a mean-field approach and third quantization
analysis, it is found that the system exhibits rich quantum phase
behaviours, including distinct superradiant fixed points, multi-phase
coexistence and limit cycle oscillation. In particular, the cavity
dissipation stabilizes a family of inverted spin coherent steady states,
whose stability region can be enlarged or reduced by properly tuning the
imbalance between the co- and counter-rotating interactions. This property
provide a conceptually new scenario to prepare coherent atomic state with
high fidelity.
\end{abstract}

\maketitle

\section{Introduction}

As the simplest model describing the coherent interaction between atomic
spins and quantized light field, the Dicke model is the key to understanding
a variety of collective phenomena in the light-matter composite system \cite%
{Dicke1,Dicke2}. The most notable prediction is the transition from a normal
phase (NP), where the photon mode is empty, to a superradiant phase (SP)
with macroscopically occupied photons and partially exited atoms \cite%
{superradiance1,superradiance2,superradiance3,superradiance4}. The
supperradiance phase transition has been observed experimentally in the
coherently driven atomic gases inside optical cavity \cite%
{DickeExperiment1,DickeExperiment2,DickeExperiment3}. An ubiquitous aspect
acquired by this system is the inevitable photon loss, which is responsible
for the dissipative evolution of dynamical variables \cite%
{open1,open2,open3,DickeTheory1,open4,DickeTheory2}. The interplay between
the coherent and dissipative dynamics in the atom-phton system may induce
novel non-equilibrium steady states, leading to intense research interest of
late on the open Dicke-like models \cite%
{Dickelike1,Dickelike2,Dickelike3,Dickelike4,Dickelike5,Dickelike6,Dickelike7,Dickelike8,Dickelike9,Dickelike10,Dickelike11,Dickelike12}%
.

In principle, the atom-photon interaction can be separated into two distinct
parts --- the \textquotedblleft co-rotating\textquotedblright\ and
\textquotedblleft counter-rotating\textquotedblright\ couplings \cite%
{book0,CRT0}. The former contains the terms which conserve excitation number
whereas the latter changes the number of excitations by two. It is the
competition between the two coupling terms, together with the contribution
from the photon dissipation, that determine the final dynamics of the system
\cite{CRT1,CRT2,CRT3}. For the standard open Dicke model, physics are frozen
to the case where the competition between the two terms is balanced, while
allowing the interaction interpolating between the \textquotedblleft
co-rotating\textquotedblright\ and \textquotedblleft
counter-rotating\textquotedblright\ dominated regimes can lead to diverse
non-equilibrium phase behaviours beyond the balanced one \cite%
{IDTC1,IDTC2,IDTC3,IDTC4,IDTC5,IDTC6,IDTC7,IDTC8}. Examples includes
multicritical points \cite{IDTC3}, limit cycles and chaotic dynamics \cite%
{IDTC1,IDTC2,IDTC5,IDTC8}, etc. Along a pioneer theoretical proposal \cite%
{DickeP}, the independent control of the \textquotedblleft
co-rotating\textquotedblright\ and \textquotedblleft
counter-rotating\textquotedblright\ interactions has been experimentally
accomplished by employing an unbalanced cavity-assisted Raman coupling in
cold atomic gases \cite{IDTCexp1,IDTCexp2,IDTCexp3}.

Another hotsopt in the realm of quantum optics is the three-level system
interacting with light, as it is related to an important class of quantum
phenomena, including electromagnetically induced transparency \cite%
{EIT1,EIT2}, lasing without inversion \cite{LWI1,LWI2}, and quantum beats in
resonance fluorescence \cite{quantumbeat1,quantumbeat2}. The extension of
the two-level Dicke model to the three-level system naturally bring about
new perspectives on the atom-photon interaction \cite%
{book0,subradiance1,ThreeDicke1,ThreeDicke2,ThreeDicke3,subradiance2,ThreeDicke4,ThreeDicke5,ThreeDicke6,ThreeDicke7,ThreeDicke8}%
, such as the time crystalline order \cite{ThreeDicke4,ThreeDicke5},
enantiodetection of chiral molecules \cite{ThreeDicke6}, and subradiance
\cite{subradiance1,subradiance2}, etc. A recently interesting finding is the
family of dark and nearly dark inverted states engineered by cavity
dissipation \cite{ThreeDicke8}. These works on the three-level Dicke model,
however, have mainly focused on atom-photon interaction with either
the\textquotedblleft co-rotating\textquotedblright\ terms only \cite%
{subradiance1,subradiance2,ThreeDicke6}, or equal \textquotedblleft
co-rotating\textquotedblright\ and \textquotedblleft
counter-rotating\textquotedblright\ couplings \cite%
{ThreeDicke4,ThreeDicke5,ThreeDicke8}, leaving the interplay between the two
coupling terms largely unexplored.

In this work, we study the system of V-typed three-level atoms interacting
with a single-mode cavity field. The cavity photons mediating the two atomic
transitions are different by a phase rotation of $\pi /2$. The model
supports independently controlled co- and counter-rotating terms, allowing
the light-matter\ interaction interpolating between different regimes.
Adopting a mean-field approach and fluctuation analysis, we provide a
systematic analysis of the quantum phase behaviour of the system. It is
found that the unbalanced light-matter coupling enriches both the closed and
open phase diagrams. The main contributions of this work is summarized as
follows.

(i) For the closed system, we find two types of superradiance phase
transitions characterized by the symmetry breaking of different $%
\mathbb{Z}
_{2}$ operations. The two superradiant phases are separated by a U(1)
symmetry line in the phase diagram. Furthermore, an excited normal phase,
coexisting with the superradiant phase, is revealed.

(ii) The dissipative nature carried by the photon leakage imposes a generic
instability on the normal phase for equal co- and counter-rotating
couplings. Away from the equal coupling case, some new steady state
behaviours, including the stabilized normal phase and a persistent
oscillatory limit cycle phase, can emerge.

(iii) The family of inverted spin coherent steady states, which is
stabilized by the cavity dissipation, is enlarged (reduced) when approaching
the counter-rotating (co-rotating) interaction side. Based on this property,
we propose a cavity-assisted atomic state preparation scenario with high
fidelity.

This work is organized as follows. In Sec.~\ref{sec:model}, we describe the
proposed model and present the Hamiltonian. In Sec.~\ref{sec:closed}, we map
out the phase diagrams for the closed system. In Sec.~\ref{sec:open}, we
show the steady-state phase diagrams for the driven-dissipative system. We
discuss the dissipation-stabilized inverted steady states and show the
related scenario to prepare coherent atomic state in Sec.~\ref{sec:inverted}
and summarize in Sec.~\ref{sec:conclusion}.
\begin{figure}[tp]
\includegraphics[width=8.5cm]{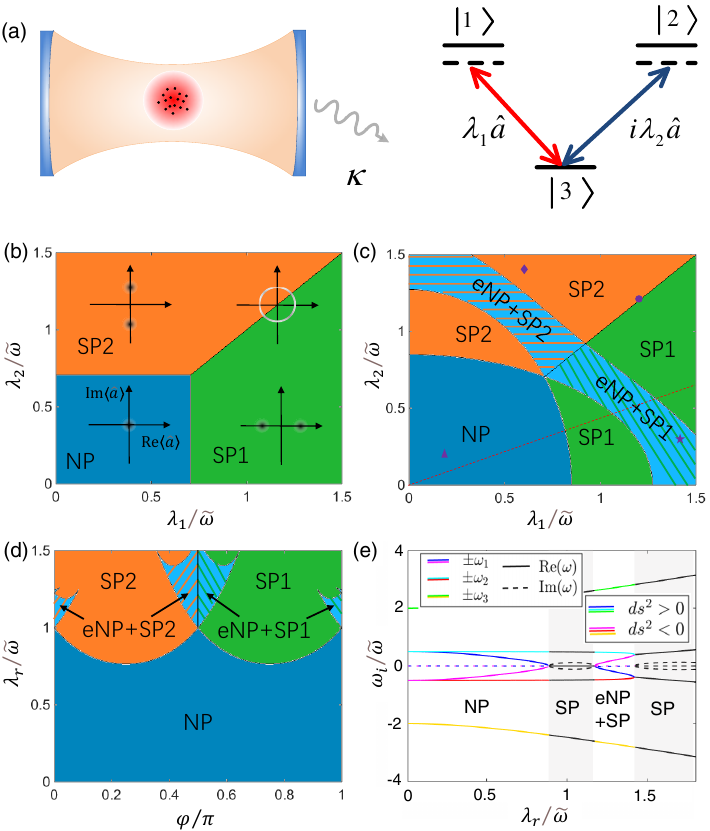}
\caption{(a) Schematic illustration of the considered setup. An ensemble of
V-typed three-level atoms are strongly coupled to a single-mode cavity field
with dissipation rate $\protect\kappa $. The cavity photons mediating the
two atomic transitions are different by a phase rotation of $\protect\pi /2$%
. (b-d) Phase diagrams of the nondissipative model showing NP (blue), SP1
(green), and\ SP2 (orange). The phase coexistence regions of e-NP and SP are
represented by different colours with hatched patterns. The phase diagrams
are plotted in the (b-c) $\protect\lambda _{1}-\protect\lambda _{2}$ plane
and (d) $\protect\varphi -\protect\lambda _{r}$ plane with (b) $\protect%
\varphi =\protect\pi /4$, (c) $\protect\varphi =7\protect\pi /16$, and (d) $%
\protect\lambda _{2}/\protect\lambda _{1}=0.41$. Four distinct phases in (a)
are indicated by their respective cavity-field distributions as a function
of the real and imaginary parts of the cavity mode Re$\left\langle \hat{a}%
\right\rangle $ and Im$\left\langle \hat{a}\right\rangle $. (e) Real (solid)
and imaginary (dashed) parts of the excitation energies $\pm \protect\omega %
_{i}$ on top of the NP along the red dotted cut line in (c). The
particlelike (holelike) fluctuations are denoted by blue, cyan, and green
(magenta, red and yellow) lines, i.e., $ds^{2}>0$ ($ds^{2}<0$).\ The
excitation spectra in the NP and e-NP are purely real whereas \ some of
their imaginary parts acquire a finite value in the SP (grey region). Note
that crossing from the NP to the e-NP, the soft-mode excitations, $\pm
\protect\omega _{1}$, swap their sign of norms, indicating a
particle-to-hole inversion. In these figures, $\protect\lambda _{r}\equiv
\protect\sqrt{\protect\lambda _{1}^{2}+\protect\lambda _{2}^{2}}$, and $%
\protect\omega =4\protect\omega _{0}=2\tilde{\protect\omega}$ with reference
frequency $\tilde{\protect\omega}\equiv \protect\sqrt{\protect\omega \protect%
\omega _{0}}$.}
\label{closed}
\end{figure}

\section{Model}

\label{sec:model}

We consider N identical V-type three-level atoms interacting with a
single-mode cavity field. Each atom consists of one lowest level $\left\vert
0\right\rangle $ and two degenerate levels $\left\vert 1\right\rangle $ and $%
\left\vert 2\right\rangle $ [see Fig.~\ref{closed}(a)]. The transitons $%
\left\vert 0\right\rangle \longleftrightarrow \left\vert 1\right\rangle $
and $\left\vert 0\right\rangle \longleftrightarrow \left\vert 2\right\rangle
$ are mediated by cavity fields with phase difference of $\pi /2$, allowing
potentially different co- and counter-rotating interactions. Such a scenario
can be effectively engineered in atomic gases with long-lived hyperfine
states. These states are then coupled by pump lasers and cavity field, which
forms, typically unbalanced, Raman transitions (see Appendix A for
descriptions of the proposed experimental configuration). The Hamiltonian
describing this system reads
\begin{eqnarray}
\hat{H} &=&\hbar \omega \hat{a}^{\dag }\hat{a}+\hbar \omega _{0}\left( \hat{%
\Lambda}_{1,1}+\hat{\Lambda}_{2,2}\right)  \notag \\
&&+\left[ \frac{\hbar \lambda _{1}}{\sqrt{N}}\hat{\Lambda}_{1,0}(\sin
(\varphi )\hat{a}+\cos (\varphi )\hat{a}^{\dag })\right.  \notag \\
&&\left. +\frac{i\hbar \lambda _{2}}{\sqrt{N}}\hat{\Lambda}_{2,0}(\sin
(\varphi )\hat{a}-\cos (\varphi )\hat{a}^{\dag })+\text{H.c.}\right] ,
\label{Ham}
\end{eqnarray}%
where $\hat{a}$ is the annihilation operator of the cavity photon, $\hat{%
\Lambda}_{i,j}=\sum_{k=1}^{N}\left\vert i\right\rangle _{k}\left\langle
j\right\vert _{k}$ ($i,j=1,2,3$) represent the collective spin operators, $%
\omega $ is the cavity frequency, $\omega _{0}$ denotes the transition
frequency between level $\left\vert 0\right\rangle $ and the two degenerate
levels $\left\vert 1\right\rangle $ and $\left\vert 2\right\rangle $, and $%
\lambda _{\mu }$ ($\mu =1,2$) are the corresponding collective coupling
strengths. Note that the parameter $\varphi $ is introduced to control the
relative weight between the co- and counter-rotating terms. Observing the
symmetry of the Hamiltonian under the transformations $\varphi \mapsto
\varphi +\pi $ and $a\mapsto -a$, we can restrict the value range of $%
\varphi $ to $[0,\pi ]$ without loss of generality. It follows that the
corotating (counter-rotating) interaction overwhelms the counter-rotating
(corotating) interaction for $\varphi \in (\pi /4,3\pi /4)$ ($\varphi \in
\lbrack 0,\pi /4)\cup (3\pi /4,\pi ]$). The pseudospin operators $\Lambda
_{i,j}$ can be\ mapped onto the Gell-Mann matrices, and thus spans the SU(3)
symmetry space of Lie algebra \cite{book2}. This is in contrast to the
pseudospin operators for the two-level Dicke model, which constitute the
SU(2) commutation relation. This difference between the two atomic
symmetries may lead to drastically different equation of motion and hence
fundamentally influence the steady states \cite%
{ThreeDicke4,ThreeDicke5,ThreeDicke8}.

Hamiltonian~(\ref{Ham}) extends the standard two-level Dicke model to
multiple distinct parameter regimes. For example, in the case of $\varphi
=\pi /4$, the transitions between the atomic lowest level and the two
excited levels are respectively coupled by two orthogonal quadratures of the
cavity field, whose non-equilibrium features was considered in Refs.~\cite%
{Dickelike9,ThreeDicke8}. While for $\lambda _{\mu }=0$ ($\mu =1,2$), the
Hamiltonian~(\ref{Ham}) reduces to the interpolating Dicke-Tavis-Cummings
model \cite{IDTC1,IDTC2,IDTC3,IDTC4,IDTC5,IDTC6,IDTC7,IDTC8}, which recovers
the standard Dicke (Tavis-Cummings) model by further setting $\varphi =\pi
/4 $ ($\varphi =\pi /2$).

In general, the Hamiltonian~(\ref{Ham}) possess a double $%
\mathbb{Z}
_{2}\otimes
\mathbb{Z}
_{2}$ symmetry, which is composed of the two other transformations $\left(
\hat{a},\hat{\Lambda}_{10},\hat{\Lambda}_{20}\right) $ $\underrightarrow{%
\mathcal{T}_{1}}$ $\left( -\hat{a}^{\dag },-\hat{\Lambda}_{01},\hat{\Lambda}%
_{02}\right) $ and $\left( \hat{a},\hat{\Lambda}_{10},\hat{\Lambda}%
_{20}\right) $ $\underrightarrow{\mathcal{T}_{2}}$ $\left( \hat{a}^{\dag },%
\hat{\Lambda}_{01},-\hat{\Lambda}_{02}\right) $. This discrete symmetry can
be enlarged to a U(1) symmetry in two specific cases: (i) $\varphi =n\pi /2$
($n\in
\mathbb{Z}
$) and (ii) $\lambda _{1}=\lambda _{2}$. Depending on the parity of $n$,
either the co-rotating or the counter-rotating term vanishes for case (i),\
leading to the U(1) symmetry found in the Tavis-Cummings (TC) model \cite{book0,TC}%
. The U(1) symmetry in case (ii) is characterized by a nontrivial
transformation $\hat{H}=$ $\hat{U}^{\dag }(\vartheta )\hat{H}\hat{U}%
(\vartheta )$ with $\hat{U}(\vartheta )=\exp (i\vartheta \hat{G})$, and $%
\hat{G}=\hat{a}^{\dag }\hat{a}+i(\hat{\Lambda}_{21}-\hat{\Lambda}_{12})$
satisfying $[\hat{H},\hat{G}]=0$. Notice that the conserved quantity $\hat{G}
$ has also been pointed out in Refs. \cite{Dickelike9,ThreeDicke8} for the
balcanced coupling case. We here show that the constraint on $\varphi (=\pi
/4)$ can be completely relaxed, yielding a continuous family of models, each
labeled by $\varphi $, that respect the same U(1) symmetry. In the spirit of
Landau's theory, the aforementioned symmetries of the Hamiltonian signals
potential equilibrium or nonequilibrium phase transitions. In the following
two sections, we provide a thorough analysis of the emergent quantum phases
for both the nondissipative and dissipative models. For each model we first
show the results of the balanced coupling case with $\varphi =\pi /4$, and
then explore the effects of deviation from this balanced point.

\section{Phase diagram for the closed system}

\label{sec:closed}

The static properties of a closed system is involved in its mean-field
energy (ME) functional, which can be formally obtained by using an SU(3)
generalization of the Holstein-Primakoff transformation (see Appendix B for
details). A fluctuation analysis around the extrema of the ME determines the
stability of various phases: the phase is physical and stable only if its
fluctuation excitations acquire a completely real spectrum.

It is found that the NP, where the cavity mode is empty and the atoms
populate the lowest level $\left\vert 0\right\rangle $, is enclosed by the
curve (Appendix C)
\begin{equation}
(2\left\vert \mathcal{B}\right\vert +L)^{2}-\omega ^{2}\omega _{0}^{2}=0,
\end{equation}%
with $L=\lambda _{1}^{2}+\lambda _{2}^{2}$ and $\mathcal{B}=\cos (\varphi
)\sin (\varphi )(\lambda _{1}^{2}-\lambda _{2}^{2})$. For parameters obeying
$(2\left\vert \mathcal{B}\right\vert +L)^{2}>\omega ^{2}\omega _{0}^{2}$,
the system enters the SP by undergoing a second-order phase transition. In
this phase, the cavity mode is macroscopically populated as $\left\vert
\left\langle a\right\rangle \right\vert $ $=\sqrt{[(2\left\vert \mathcal{B}%
\right\vert +L)^{2}-\omega ^{2}\omega _{0}^{2}]/4(2\left\vert \mathcal{B}%
\right\vert +L)\omega ^{2}}$, and the atoms are partially excited to their
higher energy levels $\left\vert 1\right\rangle $ or $\left\vert
2\right\rangle $. In the SP, the sign of $\mathcal{B}$ further\
distinguishes two distinct phases: for $\mathcal{B}>0$ ($\mathcal{B}<0$),
the cavity mode acquires a real (imaginary) macroscopic excitation with Re$%
\left\langle \hat{a}\right\rangle \neq 0$ and Im$\left\langle \hat{a}%
\right\rangle =0$ (Re$\left\langle \hat{a}\right\rangle =0$ and Im$%
\left\langle \hat{a}\right\rangle \neq 0$), and the $\mathcal{T}_{1}$ ($%
\mathcal{T}_{2}$) symmetry is spontaneously broken. We term the SP with $%
\mathcal{B}>0$ superradiant phase 1 (SP1), and that with $\mathcal{B}<0$
superradiant phase 2 (SP2).\ The critical curve $\mathcal{B}=0$, along which
the Hamlitonian respects a U(1) symmetry, determines a first order phase
boundary between the SP1 and SP2.

A typical parameters choice is the balanced driving case with $\varphi =\pi
/4$. In this case, the counter-rotating and corotating interactions feature
on an equal footing. The closed phase diagram is outlined in Fig. \ref%
{closed}(b). For $\lambda _{1},\lambda _{2}\leqslant $ $\lambda _{c}\equiv
\sqrt{\omega \omega _{0}/2}$, the system is located in the NP. Tunning one
of the coupling strength above the critical value $\lambda _{c}$, namely$\
\max (\lambda _{1}$,$\lambda _{2})>\lambda _{c}$, the system enters the SP.
The U(1)-symmetry line $\lambda _{1}=\lambda _{2}>\lambda _{c}$ splits the
SP into two subphases: the SP1 with $\lambda _{1}>\lambda _{2}$ and the SP2
with $\lambda _{2}>\lambda _{1}$.

Allowing the coupling strength of the counter-rotating and corotating terms
unbalanced, say $\varphi \neq \pi /4$, results in richer phenomena. The
phase diagram of $\varphi =7\pi /16$ is representatively plotted in Fig.~\ref%
{closed}(c). Different from the balanced case [Fig.~\ref{closed}(b)], deep
inside the SP, a considerably large region where NP is also stable, emerges.
We remark that this NP is essentially a stable exited state since it
corresponds to a local maximum of the ME landscape (see Appendix C for
detailed description). Following the nomenclature used in Ref.~\cite{IDTC6},
we hereafter dub the NP, which coexists with the SP, exited-Normal phase
(e-NP). To see the impacts of the unbalanced co- and counter-rotating
interactions more clearly, we plot in Fig.~\ref{closed}(d) the phase diagram
as a function of $\varphi $ and the coupling strength $\lambda _{r}\equiv
\sqrt{\lambda _{1}^{2}+\lambda _{2}^{2}}$ with $\lambda _{2}/\lambda
_{1}=0.41$. It is to be seen that, as the system deviates away from the
balanced point $\varphi =\pi /4$, the regions of the phase coexistence of SP
and e-NP becomes pronounced.

Apart from the ME landscape, the NP and e-NP are dynamically distinct by the
nature of excitations: at positive (negative) eigenfrequencies, the
soft-mode excitations of both NP and SP are particlelike (holelike), whereas
those of e-NP are holelike (particlelike) \cite{IDTC6,IDTC7}. This can be
confirmed by investigating the sympletic norm, $ds_{\mathbf{v}%
_{j}}^{2}\equiv $ $\mathbf{v}_{j}^{\dag }I_{z}\mathbf{v}_{j}$, defined at
each normal-mode eigenfrequency, where $\mathbf{v}_{j}$ with $j=1,...,2N$
are the eigenvectors of the Hopfeld-Bogoliubov matrix $D_{\text{H}}$ (see
Appendix C), and $I_{z}=1_{N}\otimes (-1_{N})$ \ is a $2N\times $ $2N$
diagonal matrix with $+1$ ($-1$) entries on the first (second) $N$ elements.
The nature of the excitations is intimately related to the sign of $ds_{%
\mathbf{v}_{j}}^{2}$. That is, the soft mode is a particlelike (holelike)
excitation at positive eigenfrequencies for $ds_{\mathbf{v}_{j}}^{2}>0$ ($%
ds_{\mathbf{v}_{j}}^{2}<0$), and is a holelike (particlelike) excitation at
negative eigenfrequencies for $ds_{\mathbf{v}_{j}}^{2}<0$ ($ds_{\mathbf{v}%
_{j}}^{2}>0$). Figure \ref{closed}(e) depicts the excitation spectra and the
sign of their sympletic norms on top of the NP along a representative
trajectory in parameter space [cf. Fig~\ref{closed}(c)]. As the coupling
strength increases, the system traverses NP, SP, coexistence of e-NP and SP,
and eventually end in SP. While, as expected, the whole spectra are purely
real in the NP and e-NP, the soft-mode pair, $\pm \omega _{1}$, swap their
sign of sympletic norms, indicating a particle-to-hole inversion.

\section{Steady state in the presence of cavity dissipation}

\label{sec:open}

The above picture fundamentally changes if the dissipative nature is
explicitly considered. To provide an understanding of the open phase
diagram, we start from the master equation of the form $\partial _{t}\hat{%
\rho}=\mathcal{\hat{L}}\hat{\rho}$, where the Liouvillian acts as $\mathcal{%
\hat{L}}\hat{\rho}=-i/\hbar \lbrack \hat{H},\hat{\rho}]+\kappa (2\hat{a}\hat{%
\rho}\hat{a}^{\dag }-\hat{a}^{\dag }\hat{a}\hat{\rho}-\hat{\rho}\hat{a}%
^{\dag }\hat{a})$ with $\kappa $ being the photon loss rate. The
steady-state properties of the open system are captured by a stability
analysis of the Liouvillian's fixed points, which can be effectively
achieved under the framework of third quantization \cite%
{ThirdQuant1,ThirdQuant2}. This approach produces a set of rapidities \{$%
\zeta _{i}$\}, whose role resembles that of the excitation spectrum of
closed systems: the real and imaginary parts of $\zeta _{i}$ characterizes
the lifetime and frequency of the corresponding fluctuation mode,
respectively. The steady state is stable when the real parts of all the
rapidities are nonnegative, i.e., Re$\zeta _{i}\geqslant 0$. The detailed
calculations of \{$\zeta _{i}$\} is attributed to the Appendix D. In
principle, the stable attractors of the open system can either lie in the
low energy sectors with most of the atoms populating the lowest energy level
$\left\vert 0\right\rangle $, or the high energy sectors where the atomic
population are completely inverted to the exited states $\left\vert
1\right\rangle $ and $\left\vert 2\right\rangle $. The superradiant features
can only be highlighted in the low energy sectors, which is the focus of
this Section. We leave the discussion of relevant physics in the high energy
sectors in Sec.~\ref{sec:inverted}.

\begin{figure}[tp]
\includegraphics[width=8.5cm]{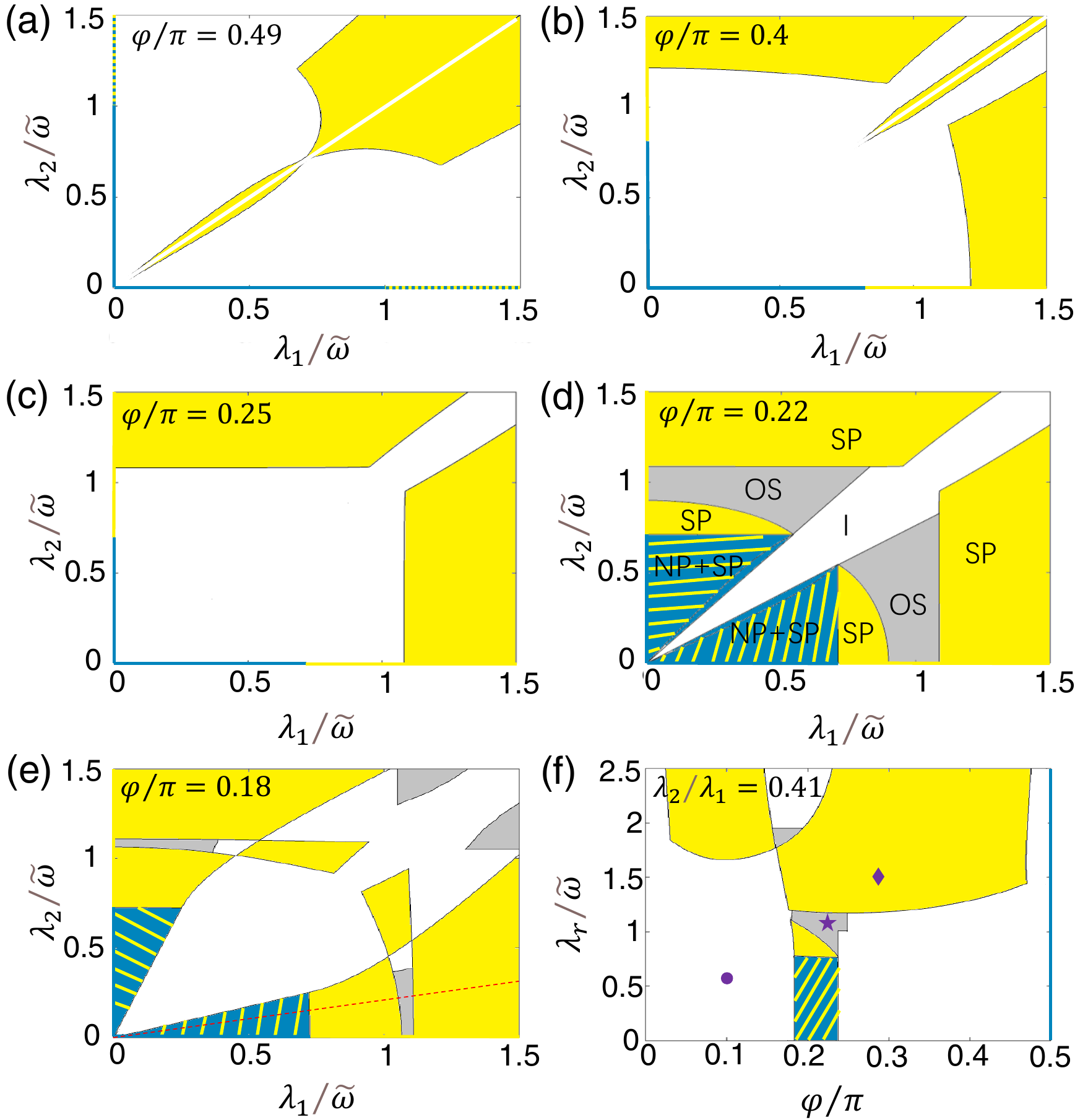}
\caption{Steady-state phase diagrams of the dissipative model showing NP
(blue), SP (yellow), OS (grey) and coexistence of NP and SP (hatched yellow
and blue) with $\protect\omega =4\protect\omega _{0}=2\tilde{\protect\omega}$
and $\protect\kappa =0.1\tilde{\protect\omega}$. Note that, except for the
two limits $\protect\lambda _{1}\cdot \protect\lambda _{2}=0$ and $\protect%
\varphi =(n+1)\protect\pi /2$ ($n\in \mathbb{Z}$), the inverted states are
stable throughout the whole parameter space. They are either the exclusive
steady states (white) or coexistent with other phases (indicated by colours
other than white). (a-e) Evolution of the phase diagram in $(\protect\lambda %
_{1}$, $\protect\lambda _{2})$ plane as one varies $\protect\varphi $ from
the co-rotating side $\protect\varphi /\protect\pi =0.49$ to the
counter-rotating side $\protect\varphi /\protect\pi =0.18$. (f) $\protect%
\lambda _{r}$ vs $\protect\varphi $ for fixed $\protect\lambda _{2}/\protect%
\lambda _{1}=0.41$.}
\label{open}
\end{figure}

A generic impact of the cavity dissipation imposed on the system is the
elimination of the U(1)-symmetry-broken phase \cite%
{IDTC3,Dickelike2,Dickelike3,Dickelike4,Dickelike9,ThreeDicke8} along the
critical curve $\mathcal{B}=0$. The SP, which features both populated real
and imaginary quadratures of the cavity mode in the open case (i.e., Re$%
\left\langle \hat{a}\right\rangle \cdot $Im$\left\langle \hat{a}%
\right\rangle \neq 0$), is stable inside multiple disconnected phase regions
separated by $\mathcal{B}=0$. As shown in Fig.~\ref{open}(c), the open phase
diagram of $\varphi =\pi /4$ is sharply different from its closed
counterpart [cf. Fig.~\ref{closed}(b)] in the following aspects: \cite%
{Dickelike9} (i) the NP is generically destabilized in the parameter space,
except for the two-level limit $\lambda _{\mu }=0$ ($\mu =1,2$), (ii) the SP
along a $\kappa $-dependent sliver around the U(1) symmetry line $\lambda
_{1}=\lambda _{2}$, vanishes, and (iii) the continuous phase boundary
enclosing the superradiant region with $\lambda _{1}\cdot \lambda _{2}\neq 0$%
, becomes first order.

The phase diagram exhibits distinctly different features in the co- and
counter-rotating-dominated regimes. We first pay attention to the corotating
side with $\varphi \in (\pi /4,\pi /2]$ \cite{remark}. For corotating
coupling strength slightly larger than the counter-rotating one, two small
islands of SP, splitted by the U(1) line $\lambda _{1}=\lambda _{2}$, emerg
inside the $\kappa $-dependent sliver [Fig.~\ref{open}(b)].\ As $\varphi $
increases further, the area of the two SP islands enlarges and even
percolates to parameter space with extremely small coupling strength $%
\lambda _{\mu }$ ($\mu =1,2$), and the $\kappa $-dependent sliver which
prevents the superradiance transition is eventually destroyed [Fig.~\ref%
{open}(a)].

\begin{figure}[tp]
\includegraphics[width=8.5cm]{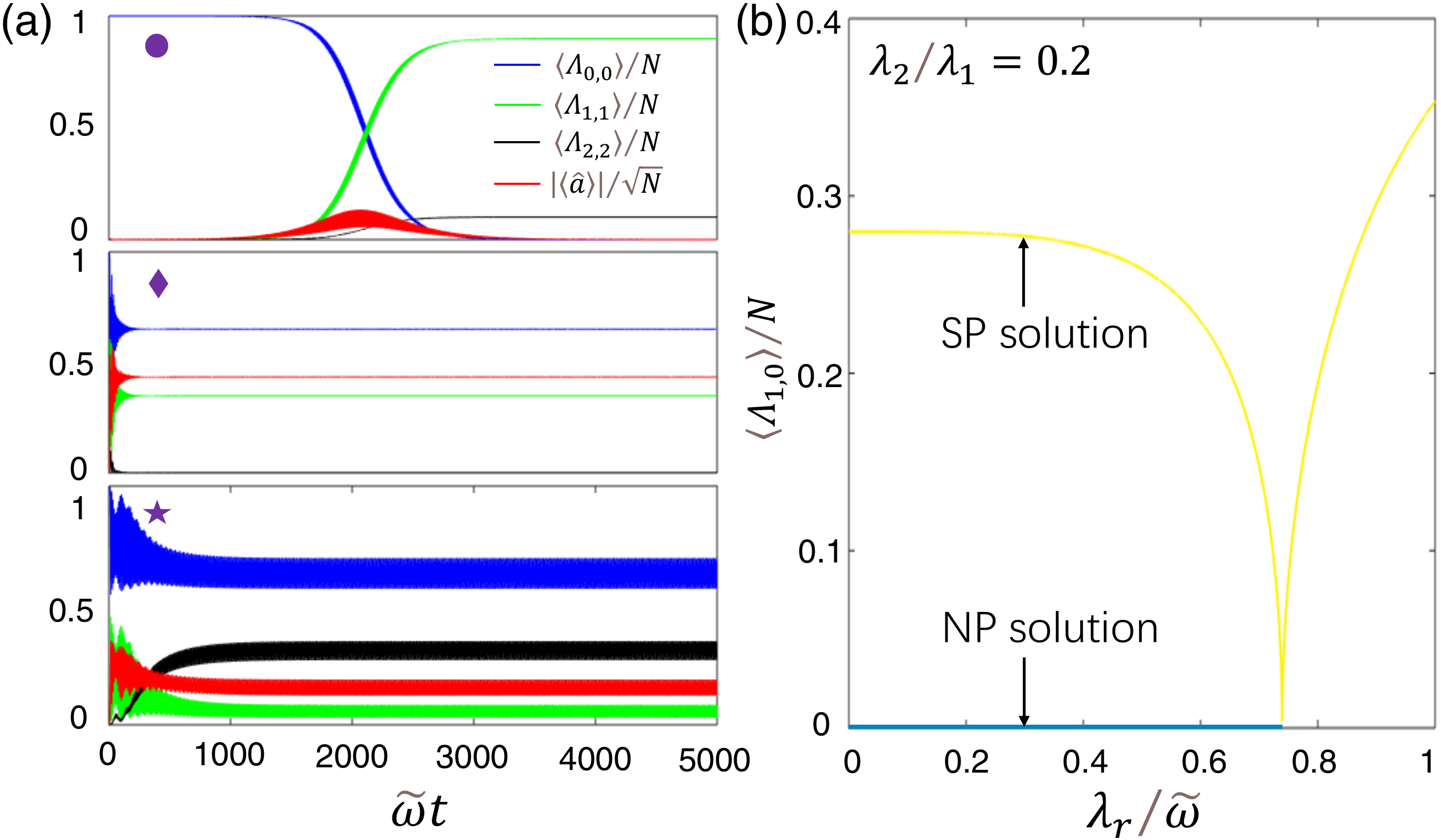}
\caption{(a) Dynamical evolutions of the cavity field amplitude and atomic
populations for the corresponding parameter locations indicated by the
symbols of round, diamond, and star in Fig.~\protect\ref{open} (f). In this
simulation, the initial state is chosen as the normal state ($\left\langle
\hat{\Lambda}_{0,0}\right\rangle =N$, $\left\langle \hat{\Lambda}%
_{0,j}\right\rangle =\left\langle \hat{\Lambda}_{i,j}\right\rangle $ for $%
i,j\in \{1,2\}$) with a small cavity field $\left\langle \hat{a}%
\right\rangle /\protect\sqrt{N}=0.01$ (see Appendix E for the equations of
motion). (b) The stable steady-state solutions of the atomic field $%
\left\langle \hat{\Lambda}_{0,1}\right\rangle $ as a function of $\protect%
\lambda _{r}$ for fixed $\protect\varphi =0.18\protect\pi $ and $\protect%
\lambda _{2}/\protect\lambda _{1}=0.2$, i.e., along the red dotted cut line
in Fig.~\protect\ref{open}(e). }
\label{open2}
\end{figure}

The physics in the counter-rotating-dominated side with $\varphi \in \lbrack
0,\pi /4)$ is richer. The first finding is the appearance of steady-state
solutions converging to limit cycles instead of fixed points, as denoted in
Figs.~\ref{open}(d)-(f). The limit cycles dictate an oscillatory
supperradiant phase (OS) in which the order parameters exhibit persistent
oscillation around some nonzero values. Figure~\ref{open2}(a) shows the
dynamical evolutions of the order parameters in three different parameter
regimes. While the steady state belonging to SP is time independent [middle
panel of Fig.~\ref{open2}(a)], the stable oscillatory character of dynamical
variables in the OS is clearly identified after a sufficiently long
integration time [bottom panel of Fig.~\ref{open2}(a)]. We remark that the
regimes of persistent oscillations also exist in the open SU(2) Dicke model
with unbalanced coupling \cite{IDTC5,IDTCexp1}. It is also found that the
NP, which is generically destabilized in the balanced coupling case, stably
coexists with the SP in a pie-chart-shaped region in $\lambda _{1}-\lambda
_{2}$ plane [Figs.~\ref{open}(d)-(e)]. In this multi-phase coexistence
region, the SP solution looks a bit counter-intuitive as it decreases to
zero as the coupling strength increases [see Fig.~\ref{open2}(b) for
illustration]. This is in sharp contrast to the standard Dicke model \cite%
{Dicke2}, where a monotonic increasing behavior of the order parameters is
observed. The critical value of the coupling strength, at which the order
parameters of the SP vanish, defines a second-order phase boundary. We
emphasize that, except for the continuous phase boundary appeared here and
those for the two-level limit $\lambda _{\mu }=0$ ($\mu =1,2$), all the
other steady-state phase transitions with $\kappa \neq 0$ are of first
order. In $\lambda _{1}-\lambda _{2}$ plane, the area of NP-SP coexisting
phase reduces as the system approaches the counter-rotating-dominated side,
until it vanishes at a critical value $\varphi _{c}$. The existence of such
criticality becomes immediately clearer if we plot the phase diagram as a
function of $\varphi $ and $\lambda _{r}$ for fixed $\lambda _{1}/\lambda
_{2}$ [Fig.~\ref{open}(f)]. As another interesting aspects demonstrated by
this figure, while the NP keeps stable with $\varphi =\pi /2$, an infinitely
small counter-rotating fraction may destabilize it and drive the fixed
points to a family of inverted states in the high energy sectors
[represented by the white region in Fig.~\ref{open}(f)]. The
counter-rotating terms represent a process explicitly breaking the energy
conservation, which is commonly believed of less significance for weak
enough coupling strength \cite{book}. Our results here show that these
terms, although vanishingly small, deserves special attention when the
atomic symmetry is enlarged. An in-depth investigation of this subject is
out of the scope of this paper and will be attributed to future work.

\section{Dissipation stabilized Inverted state}

\label{sec:inverted}

Up to now, the quantum states we discussed are restricted to the low energy
sectors where the atomic lowest level $\left\vert 0\right\rangle $ is
macroscopically populated. There is, however, a different class of states
with unoccupied $\left\vert 0\right\rangle $. These inverted states, having
a much higher energy than those of the NP and SP, are characterized by two
parameters
\begin{equation}
N_{1}=\left\langle \hat{\Lambda}_{11}\right\rangle \text{, and }\theta =\arg
\left\langle \hat{\Lambda}_{12}\right\rangle  \label{InverPara}
\end{equation}%
which respectively denotes the occupation of level $\left\vert
1\right\rangle $ and the relative phase between levels $\left\vert
1\right\rangle $ and $\left\vert 2\right\rangle $. The collective states
determined by parameters~(\ref{InverPara}) are essentially spin coherent
state in the inverted-state subspace. Of particular importance in the class
of inverted spin coherent states is the dark state defined as \cite%
{Dark1,Dark2}


\begin{equation}
\left\vert D\right\rangle =\prod_{j=1}^{N}\left\vert d\right\rangle _{j},
\label{darks}
\end{equation}%
where $\left\vert d\right\rangle =$ $i\sin (\nu )\left\vert 1\right\rangle
+\cos (\nu )\left\vert 2\right\rangle $ and $\tan (\nu )=\lambda
_{2}/\lambda _{1}$. Note that with this definition, the state $\left\vert
D\right\rangle $\ is uniquely defined by the parameter $\nu $. The dark
state~(\ref{darks}) is completely decoupled from the radiation field and
therefore becomes a stable eigenstate of the Hamiltonian~(\ref{Ham}). The
lack of adiabatic passage makes the inverted states less important in the
closed system. They, nevertheless, become crucial under the open environment
due to their accessibility provided by the cavity dissipation \cite%
{IDTC1,IDTC2,IDTC6,IDTC7}.

It should be noticed that, while all the inverted spin coherent states turn
out to be fixed points of the Liouvillian $\mathcal{\hat{L}}$, only a subset
of them is stable. By analyzing the related rapidities, it is found that the
stable fixed points fall into a region enclosed by a stability boundary in
the $\theta -N_{1}$ plane,
\begin{equation}
\frac{2\eta _{1}\eta _{2}\sin (\theta )}{\eta _{1}^{2}+\eta _{2}^{2}}=\frac{%
(\kappa ^{2}+\omega ^{2}+\omega _{0}^{2})\cos (2\varphi )-2\omega _{0}\omega
}{\kappa ^{2}+\omega ^{2}+\omega _{0}^{2}-2\omega _{0}\omega \cos (2\varphi )%
}\equiv \Omega  \label{boundary}
\end{equation}%
where $\eta _{1}=$\ $\lambda _{1}N_{1}/\sqrt{N}$ and $\eta _{2}=\lambda _{2}%
\sqrt{1-N_{1}/N}$, and the role of the parameter $\varphi $ is encapsulated
in the scaled variable $\Omega $. Setting $\varphi =\pi /4$, we reproduce
the result of the balanced case obtained in Ref. \cite{ThreeDicke8}, in
which the value of $\Omega $ are restricted in between $0$ and $1$ by
definition. Allowing the parameter $\varphi $ tunable, however, feasible
range of the scaled variable $\Omega $ is extended to $[-1,1]$. As is
detailed in the following, the enlargement of the value range of $\Omega $
provides new possibilities to engineer the atomic steady state.

\begin{figure}[tp]
\includegraphics[width=8.5cm]{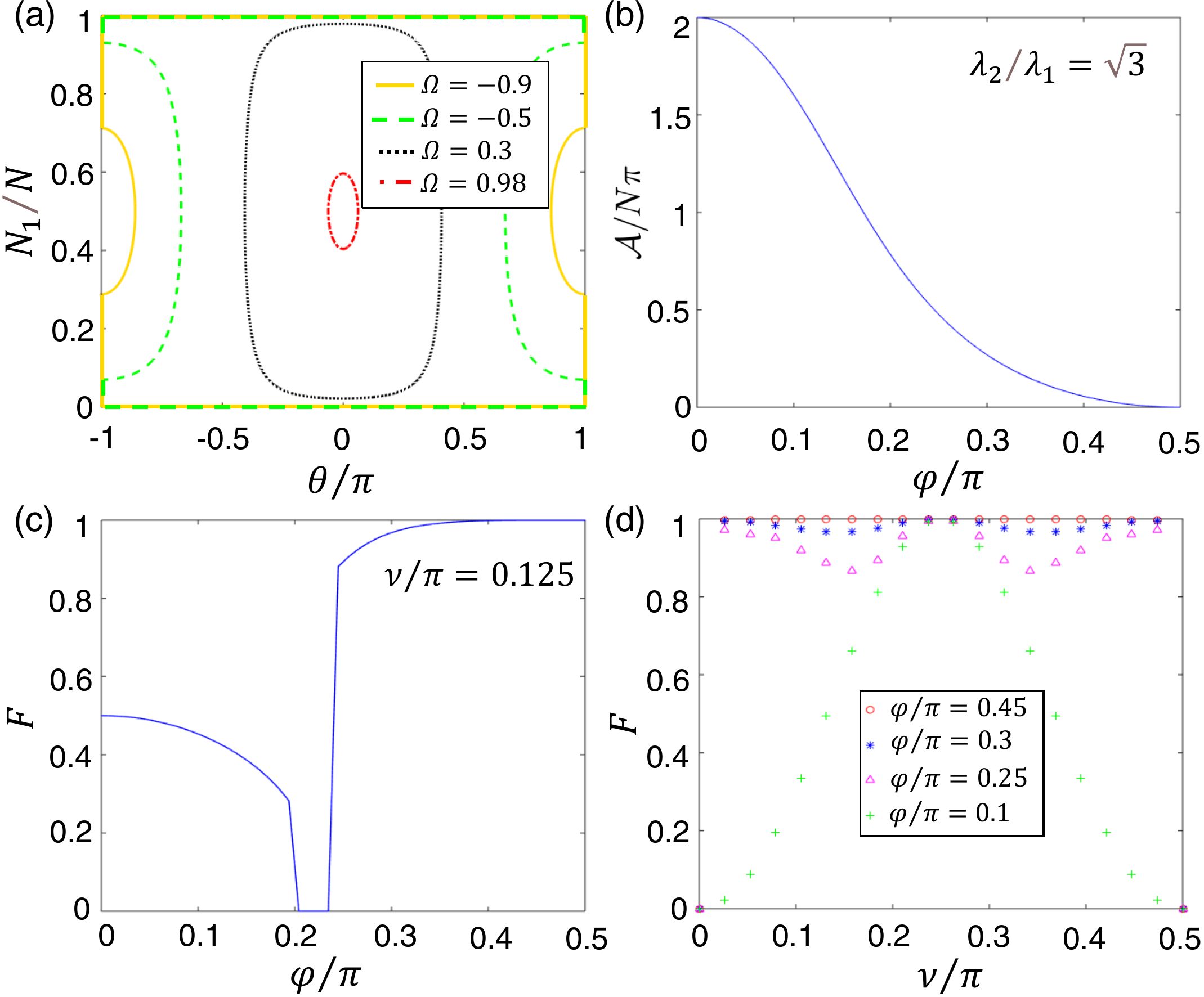}
\caption{(a) Stability boundaries of the inverted state in the $\protect%
\theta -N_{1}$ parameters space for $\protect\lambda _{1}=\protect\lambda %
_{2}$ and different $\Omega $. Solutions inside the regions enclosed are
stable. Note that here the parameter $\protect\theta $ is shifted by $%
\protect\pi /2$ for clarity. (b) Evolution of the area $\mathcal{A}$ as a
function of $\protect\varphi $ for $\protect\lambda _{2}/\protect\lambda %
_{1}=\protect\sqrt{3}$ and $\protect\kappa =0.1\tilde{\protect\omega}$.
(c-d) Fidelities given by Eq.~(\protect\ref{FID}) for $\protect\kappa =%
\tilde{\protect\omega}$ as a function of (c) $\protect\varphi $ with $%
\protect\nu =\protect\pi /8$, and (d) $\protect\nu $ with varying $\protect%
\varphi $. The density matrix $\protect\rho _{s}$ is obtained by \
integrating the mean-field equations of motion until a steady state can be
identified. The other parameters are $\protect\omega =4\protect\omega _{0}=2%
\tilde{\protect\omega}$.}
\label{Dark}
\end{figure}

With the stability boundary defined in Eq.~(\ref{boundary}), the area of the
enclosed region is derived as $\mathcal{A}=$ $N\pi \lbrack 1-\Omega (\lambda
_{1}^{2}+\lambda _{2}^{2})/\sqrt{\Omega ^{2}(\lambda _{1}^{2}-\lambda
_{2}^{2})^{2}+4\lambda _{1}^{2}\lambda _{2}^{2}}]$ \cite{ThreeDicke8}. We
plot the the stability boundary Eq.~(\ref{boundary}) for several
representative parameters in Fig.~\ref{Dark}(a). The trend that $\mathcal{A}$
increases with the decrease of $\Omega $ is obvious. More importantly,
tuning the model from the corotating side to the counter-rotating side by
varying $\varphi $, $\Omega $ goes from $1$ to $-1$, and the area of the
stable region increases from $0$ to $2\pi N$, as is illustrated in Fig.~\ref%
{Dark}(b).

The two end points, $\varphi =0$ and $\varphi =\pi /2$, deserve special
attention, since the values of $\mathcal{A}$ in these two cases\ are best
understandable through a purely physical argument. The multistability for a
more general value $\varphi \in (0,\pi /2)$ interpolate between the two
limit cases. Let us focus on $\varphi =\pi /2$ first. In this case, the
Hamiltonian~(\ref{Ham}) is simplified to a three-level TC-like model,
\begin{equation}
\hat{H}=\hat{H}_{TC}+\omega _{0}\hat{\Lambda}_{d,d},
\end{equation}%
where $\hat{H}_{TC}=$ $\hbar \omega \hat{a}^{\dag }\hat{a}+\omega _{0}\hat{%
\Lambda}_{r,r}+\hbar \lambda _{r}(\hat{\Lambda}_{r,0}a+\hat{\Lambda}%
_{0,r}a^{\dag })$ is the TC Hamiltonian with $\left\vert r\right\rangle
=\cos (\nu )\left\vert 1\right\rangle +i\sin (\nu )\left\vert 2\right\rangle
$ being\ a single-particle bright state. Since the open TC model stablizes
only the NP \cite{IDTC3}, the dark state $\left\vert D\right\rangle $
becomes the only stable inverted state, manifesting a single point in the $%
\theta -N_{1}$ parameters space (i.e., $\mathcal{A}=0$). We then turn to the
other limit $\varphi =0$, where the light-matter interaction is purely
governed by the counter-rotating terms. In this regime, the variation of the
excitation number for both spin and bosonic parts is one, whereas that for
the light-matter polariton mode is two. It is straightforward to show that
this scheme prohibits the direct transition between the atomic inverted
states with vacuum photon mode and any other states in the Hilbert space.
Hence, atoms in the inverted-state subspace are all decoupled from the
radiation field, meaning that the stable region occupies the whole $\theta
-N_{1}$ parameters space.

The preparation of a spin coherent state with required population projection
on the levels $\left\vert 1\right\rangle $ and $\left\vert 2\right\rangle $
is always one of the central aims in the atomic physics. The fact that our
model hosts a single stable inverted state $\left\vert D\right\rangle $ in
the TC limit $\varphi =\pi /2$, for which the population projection is tuned
by the system parameter $\nu $, suggests a potential scenario to achieve
this goal. However, in this case, a general initial state will not evolve to
the state $\left\vert D\right\rangle $ due to its\ darkness, but instead
dissipates to the lowest level $\left\vert 0\right\rangle $ \cite{CRT0}.
Fortunately, as is illustrated in Sec.~\ref{sec:open}, even an infinitely
small counter-rotating coupling can destabilize the level $\left\vert
0\right\rangle $ and drives the fixed points to a multistable region of
inverted states, which are bounded by a closed curve in the $\theta -N_{1}$
plane. Considering the region of multistability shrinks to the
representative point of $\left\vert D\right\rangle $ as the counter-rotating
coupling strength decreases to zero, a natural anticipation is, from a
general initial state, the steady state can approache the dark state $%
\left\vert D\right\rangle $ in a similar fashion. To verify this, we can
look at the fidelity of the steady state \cite{Fidelity1,Fidelity2},
\begin{equation}
F=Tr\left( \hat{\rho}_{s}\hat{\rho}_{d}\right)  \label{FID}
\end{equation}%
where $\hat{\rho}_{d}$ and $\hat{\rho}_{s}$ denote the density operators of
the dark state $\left\vert D\right\rangle $, which can be the target state
in demand, and the steady state of the master equation $\partial _{t}\hat{%
\rho}=\mathcal{\hat{L}}\hat{\rho}$, respectively. The fidelity Eq.~(\ref{FID}%
) quantifies the similarity between $\hat{\rho}_{s}$ and $\hat{\rho}_{d}$,
and it\ turns out to be 1 if $\hat{\rho}_{s}=\hat{\rho}_{d}$, otherwise $%
0\leqslant F<1$. Figure.~\ref{Dark}(c) depicts the fidelity $F$ as a
function of $\varphi $ for fixed $\lambda _{1}$, $\lambda _{2}$ and $\kappa $%
. As expected, in the co-rotating dominated regime, the fidelity increases
and finally approaches identity as $\varphi $ gets close to $\pi /2$. Note
that $F$ touches zero in an intermediate region, due to the stabilization of
the NP by the counter-rotating interaction [cf. Fig.~\ref{open}(f)]. To
demonstrate the feasibility of preparing a steady state with arbitrary
population projection on the levels $\left\vert 1\right\rangle $ and $%
\left\vert 2\right\rangle $, we plot $F$ as a function of $\nu $ for varying
$\varphi $ in Fig.~\ref{Dark}(d). It can be seen clearly that, tuning $%
\varphi $ to the co-rotating-dominated side, $F$ gets closed to some
constant for all values of $\nu \in (0,\pi /2)$, despite of small
fluctuations. More importantly, the closer to the co-rotating side, the
higher the fidelity is.

Before ending this section, we make two remarks. Firstly, the above
predictions for the fidelity depend crucially on the SU(3) atomic symmetry.
The vanishing of either the coupling strength $\lambda _{1}$ or $\lambda
_{2} $ reduces the atomic symmetry to SU(2), and thus essentially changes
the system dynamics. This explains the two exceptions occurring for $\nu =0$
and $\pi /2$ in Fig.~\ref{Dark}(d), where $F$ drops to zero. Secondly, the
proposed approach of state preparation can make the population projections
on levels $\left\vert 1\right\rangle $ and $\left\vert 2\right\rangle $
controllable,\ but leaves the relative phase between them fixed. The
engineering of an inverted steady state with arbitrary relative phase can be
achieved by encoding a tunable phase-difference rotation between the cavity
photons which mediate the two atomic transitions $\left\vert 0\right\rangle
\longleftrightarrow \left\vert 1\right\rangle $ and $\left\vert
0\right\rangle \longleftrightarrow \left\vert 2\right\rangle $. An in-depth
investigation of this scenario, albeit at the cost of added complexity,
merits a separate work.

\section{Conclusions}

\label{sec:conclusion}

We have investigated a system of V-typed three-level atoms interacting with
a single-mode cavity field, with main focus on the consequences of the
competition between the co- and counter-rotating interaction. Using a
mean-field approach and third quantization analysis, we have mapped out the
phase diagram of the system for both closed and open conditions. Rich
quantum phase behaviours, including multi-phase coexistence and limit cycle
oscillation, has been revealed. Of particular interesting is the inverted
spin coherent steady states stabilized by the cavity dissipation. By
analyzing the roles of the co- and counter-rotating terms in the
inverted-state stabilization, we have proposed a high-fidelity state
preparation scenario.

\acknowledgments

This work is supported partly by the National Key R\&D Program of China
under Grant No.~2017YFA0304203; the NSFC under Grants No.~12174233 and
No.~11804204.

\vbox{\vskip1cm} \appendix

\section{Effective Hamiltonian and proposed experimental implementation}

\begin{figure*}[tp]
\includegraphics[width=17.0cm]{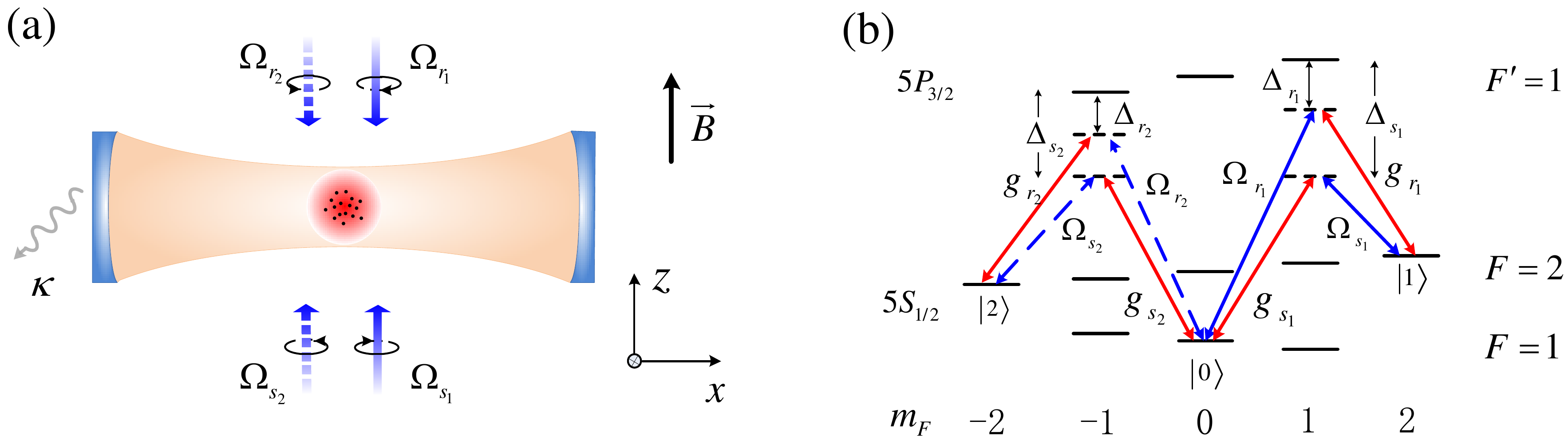}
\caption{(a)The proposed experimental setup and (b) possible atomic
excitation scheme based on the $D_{2}$ line of $^{87}$Rb atom.}
\label{setup}
\end{figure*}

In this Section, we propose an experimental implementation of our model
based on\ distinct cavity-assisted Raman transitions of cold atoms \cite%
{DickeP}. As shown in Fig.~\ref{setup}(a), an ensemble of $^{87}$Rb atoms is
trapped within an optical cavity by an intracavity optical lattice \cite%
{IDTCexp1,IDTCexp2}. The atoms are driven transverse to the cavity by two
pairs of lasers. Each pair of the lasers is composed of two
counter-propagating single beams with different circular polarization. A
guided magnetic field $B$ is applied along the direction of the laser
propagation ($z$ direction) to fix a quantized axis and split the Zeeman
sublevels of the atomic ensemble, which confirms the distinct Raman
channels. The cavity field is linearly polarized along the $y$ axis, which
is perpendicular to the magnetic field. The three hyperfine sublevels of 5$%
S_{1/2}$, $\left\vert F=2,m_{F}=+2\right\rangle $, $\left\vert
F=1,m_{F}=0\right\rangle $, and $\left\vert F=2,m_{F}=-2\right\rangle $,
can\ play the roles of atomic levels $\left\vert 1\right\rangle ,$ $%
\left\vert 0\right\rangle ,$ and $\left\vert 2\right\rangle $, respectively.
The two pairs of counter-propagating lasers, with Rabi frequencies (phases) $%
\Omega _{s_{1,2}}$ and $\Omega _{r_{1,2}}$ ($\theta _{s_{1,2}}$ and $\theta
_{r_{1,2}}$), provide optical couplings between 5$S_{1/2}$ and 5$P_{3/2}$,
and thus\ form four distinct Raman transitions, as shown in Fig.~\ref{setup}%
(b). The detunnings of driving lasers from the excited states, $\Delta
_{s_{1,2}}$ and $\Delta _{r_{1,2}}$, are assumed large enough so that we can
adiabatically eliminate the 5$P_{3/2}$ levels, yielding an effective
Hamiltonian

\begin{widetext}
\begin{eqnarray}
\hat{H} &=&\omega _{A}\hat{a}^{\dagger }\hat{a}+\sum_{j=1}^{N}\left( \omega
_{00}\left\vert 0\right\rangle _{j}\left\langle 0\right\vert _{j}+\omega
_{10}\left\vert 1\right\rangle _{j}\left\langle 1\right\vert _{j}+\omega
_{20}\left\vert 2\right\rangle _{j}\left\langle 2\right\vert _{j}\right)
\notag \\
&&+\hat{a}^{\dagger }\hat{a}\sum_{j=1}^{N}\left[ g_{r}(\mathbf{r}_{j})\left(
\frac{\left\vert 1\right\rangle _{j}\left\langle 1\right\vert _{j}}{\Delta
_{r_{1}}}+\frac{\left\vert 2\right\rangle _{j}\left\langle 2\right\vert _{j}%
}{\Delta _{r_{2}}}\right) +g_{s}(\mathbf{r}_{j})\left\vert 0\right\rangle
_{j}\left\langle 0\right\vert _{j}\left( \frac{1}{\Delta _{s_{1}}}+\frac{1}{%
\Delta _{s_{2}}}\right) \right]   \notag \\
&&+\sum_{\tau =1}^{2}\sum_{j=1}^{N}\left[ \frac{\Omega _{s_{\tau }}g_{s}(%
\mathbf{r}_{j})}{2\Delta _{s_{\tau }}}\hat{a}\left\vert \tau \right\rangle
_{j}\left\langle 0\right\vert _{j}e^{-i(\mathbf{k}_{s_{\tau }}\cdot \mathbf{r%
}_{j}+\theta _{s_{\tau }})}+\frac{\Omega _{r_{\tau }}g_{r}(\mathbf{r}_{j})}{%
2\Delta _{r_{\tau }}}\hat{a}\left\vert 0\right\rangle _{j}\left\langle \tau
\right\vert _{j}e^{-i(\mathbf{k}_{r_{\tau }}\cdot \mathbf{r}_{j}+\theta
_{r_{\tau }})}+\text{H.c.}\right]   \label{HA1}
\end{eqnarray}%
\end{widetext}

where the definitions%
\begin{equation}
\left\vert 1\right\rangle \equiv \left\vert F=2,m_{F}=+2\right\rangle ,
\end{equation}%
\begin{equation}
\left\vert 0\right\rangle \equiv \left\vert F=1,m_{F}=0\right\rangle ,
\end{equation}%
and%
\begin{equation}
\left\vert 2\right\rangle \equiv \left\vert F=2,m_{F}=-2\right\rangle
\end{equation}%
has been made. In the Hamiltonian~(\ref{HA1}), $g_{s}(\mathbf{r}_{j})$ [$%
g_{r}(\mathbf{r}_{j})$] is the single-photon coupling strength at position $%
\mathbf{r}_{j}$ mediating the transitions $\left\vert 0\right\rangle
\longleftrightarrow \left\vert F^{\prime }=1,m_{F^{\prime }}=+1\right\rangle
$ and $\left\vert 0\right\rangle \longleftrightarrow \left\vert F^{\prime
}=1,m_{F^{\prime }}=-1\right\rangle $ ($\left\vert 2\right\rangle
\longleftrightarrow \left\vert F^{\prime }=1,m_{F^{\prime }}=-1\right\rangle
$ and $\left\vert 1\right\rangle \longleftrightarrow \left\vert F^{\prime
}=1,m_{F^{\prime }}=+1\right\rangle $), and the model paremeters $\omega
_{A} $, $\omega _{00}$, $\omega _{10}$, and $\omega _{20}$ are given by
\begin{equation}
\omega _{A}=\omega _{c}-\frac{\omega _{r_{1}}+\omega _{s_{1}}}{2}
\end{equation}%
\begin{equation}
\omega _{00}=\Omega _{r_{1}}^{2}/4\Delta _{r_{1}}+\Omega
_{r_{2}}^{2}/4\Delta _{r_{2}}
\end{equation}%
\begin{equation}
\omega _{10}=\omega _{G_{1}}+\Omega _{s_{1}}^{2}/4\Delta _{s_{1}}-\frac{%
\omega _{r_{1}}-\omega _{s_{1}}}{2}
\end{equation}%
\begin{equation}
\omega _{20}=\omega _{G_{2}}+\Omega _{s_{2}}^{2}/4\Delta _{s_{2}}-\frac{%
\omega _{r_{2}}-\omega _{s_{2}}}{2}
\end{equation}%
where $\omega _{r_{1,2}}$ and $\omega _{s_{1,2}}$ are the frequencies of the
driving lasers, $\omega _{G_{1}}$ ($\omega _{G_{2}}$) characterizes the
energy of atomic level $\left\vert 1\right\rangle $ ($\left\vert
2\right\rangle $), and $\omega _{c}$ denotes the cavity frequency. We
further assume the atoms are trapped to antinodes of the cavity field, so
that the single-photon coupling strengths can be approximately constant and
written in a position-independent form $g_{r/s}(\mathbf{r}_{j})=g_{r/s}$.
For cold atoms with temperature closed to zero, the motional effect can be
neglected meaning that the atom positions can be treated as classical
variables \cite{DickeTheory1,IDTCexp2}. Bearing these assumptions in mind,
and applying the unitary transformation
\begin{widetext}
\begin{equation}
U_{\mathbf{k}}=\prod_{j=1}^{N}\exp \left[ i(\mathbf{k\cdot r}_{j}+\frac{%
\theta _{r_{1}}+\theta _{r_{2}}-\theta _{s_{1}}-\theta _{s_{2}}}{2}%
)(\left\vert 1\right\rangle _{j}\left\langle 1\right\vert _{j}+\left\vert
2\right\rangle _{j}\left\langle 2\right\vert _{j})-i\frac{\theta
_{r_{1}}+\theta _{s_{1}}}{2}\hat{a}^{\dagger }\hat{a}\right] ,
\end{equation}%
\end{widetext}

the Hamiltonian~(\ref{HA1}) becomes
\begin{eqnarray}
H &=&\omega _{A}\hat{a}^{\dagger }\hat{a}+\omega _{00}\hat{\Lambda}%
_{0,0}+\omega _{10}\hat{\Lambda}_{1,1}+\omega _{20}\hat{\Lambda}_{2,2}
\notag \\
&&+a^{\dagger }a\left[ \hat{\Lambda}_{1,1}\frac{g_{r}}{\Delta _{r_{1}}}+\hat{%
\Lambda}_{2,2}\frac{g_{r}}{\Delta _{r_{2}}}+\hat{\Lambda}_{0,0}\left( \frac{%
g_{s}}{\Delta _{s_{1}}}+\frac{g_{s}}{\Delta _{s_{2}}}\right) \right]  \notag
\\
&&+\left[ \frac{\Omega _{s_{1}}g_{s}}{2\Delta _{s_{1}}}\hat{a}\hat{\Lambda}%
_{1,0}+\frac{\Omega _{r_{1}}g_{r}}{2\Delta _{r_{1}}}\hat{a}\hat{\Lambda}%
_{0,1}+\frac{\Omega _{s_{2}}g_{s}}{2\Delta _{s_{2}}}\hat{a}\hat{\Lambda}%
_{2,0}e^{i\tilde{\theta}}\right.  \notag \\
&&\left. +\frac{\Omega _{r_{2}}g_{r}}{2\Delta _{r_{2}}}\hat{a}\hat{\Lambda}%
_{0,2}e^{i\tilde{\theta}}+\text{H.c.}\right]  \label{HA2}
\end{eqnarray}%
where $\tilde{\theta}=(\theta _{r_{1}}+\theta _{s_{1}}-\theta
_{r_{2}}-\theta _{s_{2}})/2$ and the relation $\mathbf{k}_{r}\approx -%
\mathbf{k}_{s}=\mathbf{k}$ has been used.

When the parameters are chosen as $g_{r}/\Delta _{r_{1}}=g_{r}/\Delta
_{r_{2}}=g_{s}/\Delta _{s_{1}}+g_{s}/\Delta _{s_{2}}$ and $\tilde{\theta}%
=\pi /2$, the Hamiltonian~(\ref{HA2}) reduces to%
\begin{eqnarray}
H &=&\hbar \omega \hat{a}^{\dag }\hat{a}+\hbar \omega _{10}\hat{\Lambda}%
_{1,1}+\hbar \omega _{20}\hat{\Lambda}_{2,2}  \notag \\
&&+\left[ \frac{\hbar \lambda _{1,s}}{\sqrt{N}}\hat{a}\hat{\Lambda}_{1,0}+%
\frac{\hbar \lambda _{1,r}}{\sqrt{N}}\hat{a}\hat{\Lambda}_{0,1}+i\frac{\hbar
\lambda _{2,s}}{\sqrt{N}}\hat{a}\hat{\Lambda}_{2,0}\right.  \notag \\
&&\left. +i\frac{\hbar \lambda _{2,r}}{\sqrt{N}}\hat{a}\hat{\Lambda}_{0,2}+%
\text{H.c.}\right]  \label{HA3}
\end{eqnarray}%
where
\begin{equation}
\omega =\omega _{A}+\frac{3Ng_{r}}{\Delta _{r_{1}}}
\end{equation}%
\begin{equation}
\lambda _{\tau ,s}=\frac{\sqrt{N}\Omega _{s_{\tau }}g_{s}}{2\Delta _{s_{\tau
}}}\text{ \ (}\tau =1,2\text{)}
\end{equation}%
and%
\begin{equation}
\lambda _{\tau ,r}=\frac{\sqrt{N}\Omega _{r_{\tau }}g_{r}}{2\Delta _{r_{\tau
}}}\text{ \ (}\tau =1,2\text{)}
\end{equation}%
By requiring $\omega _{10}=$ $\omega _{20}=\omega _{0}$ and reparametrizing
the collective coupling strength as
\begin{equation}
\lambda _{\tau ,s}=\sin (\varphi )\lambda _{\tau }\text{, }\lambda _{\tau
,r}=\cos (\varphi )\lambda _{\tau }\text{ \ \ (}\tau =1,2\text{),}
\end{equation}%
Eq.~(\ref{HA3}) reduces to Hamiltonian~(\ref{Ham}) in the main text.

Based on the energy levels and their transitions of $^{87}$Rb atoms \cite%
{RbTab}, together with the current experimental conditions \cite%
{IDTCexp1,IDTCexp2}, the atom-photon coupling strength can reach $g_{r}/2\pi
=0.25$ MHz and $g_{s}/2\pi =0.14$ MHz, respectively. The number of trapped
atoms, typically $N$ $\symbol{126}$ 10$^{6}$ \cite{IDTCexp2}, appears to be
practical. The atomic detunnings $\Delta _{s_{1,2}}$ and $\Delta _{r_{1,2}}$
can range from $1$ to $100$ GHz, and the parameters ($\left\vert \Omega
_{r_{1,2}}\right\vert ,\left\vert \Omega _{s_{1,2}}\right\vert ,\kappa $)
are on the order of a few megahertz. Therefore, the condition for the
adiabatic elimination of the atomic levels, ($\left\vert \Delta
_{s_{1,2}}\right\vert ,\left\vert \Delta _{r_{1,2}}\right\vert $) $\gg $ ($%
\left\vert \Omega _{r_{1,2}}\right\vert ,\left\vert \Omega
_{s_{1,2}}\right\vert ,g_{r},g_{s}$), is well satisfied. With these
parameter setting, the collective coupling strength $\lambda _{1}$ and $%
\lambda _{2}$ can be tuned from zero to the order of megahertz, making the
superradiant condition $\lambda _{1}(\lambda _{2})\geqslant \lambda _{c}$
achievable.

\section{Holstein-Primakoff transformation and the fluctuation Hamiltonian}

In this Section, we derive the effective Hamiltonians describing
fluctuations around various quantum states. These fluctuation Hamiltonians
are necessary in analyzing the stability of considered states, and can be
formally obtained using a generalized Holstein-Primakoff transformation \cite%
{HPP1,HPP2}. For system with three atomic levels, the Holstein-Primakoff
transformation is implemented by rewriting the atomic operators $\hat{\Lambda%
}_{i,j}$ as
\begin{eqnarray}
\hat{\Lambda}_{m,m} &=&N-\sum_{j\neq m}\hat{b}_{j}^{\dag }\hat{b}_{j}\text{,
}\hat{\Lambda}_{sk}=\hat{b}_{s}^{\dag }\hat{b}_{k}\text{ (}s,k\neq m\text{)},%
\text{ }  \label{HP1} \\
\hat{\Lambda}_{s,m} &=&\hat{b}_{s}^{\dag }\sqrt{N-\sum_{j\neq m}\hat{b}%
_{j}^{\dag }\hat{b}_{j}}  \notag
\end{eqnarray}%
\ \ \ where $\hat{b}_{j}^{\dag }$ and $\hat{b}_{j}$ are bosonic creation and
annihilation operators, respectively. In Eq.~(\ref{HP1}), the subscript $m$
labels a reference state around which the fluctuations are considered. We
choose $\left\vert m\right\rangle =\left\vert 0\right\rangle $ for the
normal and superradiant states, and $\left\vert m\right\rangle =\left\vert
1\right\rangle $ for the inverted state. Employing the transformations Eq.~(%
\ref{HP1}) and choosing appropriate reference states, the Hamiltonian~(\ref%
{Ham}) can be rewritten as%
\begin{eqnarray}
\hat{H} &=&\hbar \omega \hat{a}^{\dag }\hat{a}+\hbar \omega _{0}\left( \hat{b%
}_{1}^{\dag }\hat{b}_{1}+\hat{b}_{2}^{\dag }\hat{b}_{2}\right)  \notag \\
&&+\left[ \frac{\hbar \lambda _{1}}{\sqrt{N}}\hat{b}_{1}^{\dag }\sqrt{N-\hat{%
b}_{1}^{\dag }\hat{b}_{1}-\hat{b}_{2}^{\dag }\hat{b}_{2}}(\sin (\varphi )%
\hat{a}+\cos (\varphi )\hat{a}^{\dag })\right.  \notag \\
&&+\frac{i\hbar \lambda _{2}}{\sqrt{N}}\hat{b}_{2}^{\dag }\sqrt{N-\hat{b}%
_{1}^{\dag }\hat{b}_{1}-\hat{b}_{2}^{\dag }\hat{b}_{2}}(\sin (\varphi )\hat{a%
}-\cos (\varphi )\hat{a}^{\dag })  \notag \\
&&\left. +\text{H.c.}\right] ,  \label{HS}
\end{eqnarray}%
for NP and SP, and%
\begin{eqnarray}
H &=&\hbar \omega \hat{a}^{\dag }\hat{a}+\hbar \omega _{0}N-\hbar \omega _{0}%
\hat{b}_{0}^{\dag }\hat{b}_{0}  \notag \\
&&+\left[ \frac{\hbar \lambda _{1}}{\sqrt{N}}\sqrt{N-\hat{b}_{0}^{\dag }\hat{%
b}_{0}-\hat{b}_{2}^{\dag }\hat{b}_{2}}\hat{b}_{0}(\sin (\varphi )\hat{a}%
+\cos (\varphi )\hat{a}^{\dag })\right.  \notag \\
&&\left. +\frac{i\hbar \lambda _{2}}{\sqrt{N}}\hat{b}_{2}^{\dag }\hat{b}%
_{0}(\sin (\varphi )\hat{a}-\cos (\varphi )\hat{a}^{\dag })+\text{H.c.}%
\right] ,  \label{HI}
\end{eqnarray}%
for inverted state. To facilitate the following stability analysis, the
bosonic operators are assumed to be composed of their expectation value and
a fluctuation operator, i.e.,%
\begin{equation}
\text{NP/SP}\text{: \ }\hat{a}=\sqrt{N}\alpha +\hat{c},\text{ }\hat{b}_{1,2}=%
\sqrt{N}\beta _{1,2}+\hat{d}_{1,2}  \label{BS}
\end{equation}%
\begin{equation}
\text{Inverted state}\text{: }\hat{a}=\hat{c},\text{ }\hat{b}_{0}=\hat{d}_{0}%
\text{, }\hat{b}_{2}=\hat{d}_{2}+\sqrt{N-N_{1}}e^{i\theta },  \label{BI}
\end{equation}%
where $\alpha $, $\beta _{1,2}$ and $\sqrt{N-N_{1}}e^{i\theta }$ are
expectation values to be determined by mean-field approach. Note that by
definition, the expectation values $\left\langle a\right\rangle $ and $%
\left\langle b_{0}\right\rangle $ for the inverted state are zero.
Substituting Eqs.~(\ref{BS})-(\ref{BI}) into the Hamiltonians~(\ref{HS}) and
(\ref{HI}), respectively, and doing the expansion in $1/N$, we formally
obtain%
\begin{equation}
\hat{H}=Nh_{0}+\sqrt{N}\hat{h}_{1}+\hat{h}_{2}+...  \label{HamE}
\end{equation}%
where the first term on the right hand side of Eq.~(\ref{HamE}) denotes the
ME,%
\begin{equation}
E=Nh_{0}=N\hbar \omega \left\vert \alpha \right\vert ^{2}-N\hbar (r\alpha
^{\ast }+r^{\ast }\alpha )\sqrt{k}-Nk\hbar \omega _{0}+N\hbar \omega _{0}
\label{Me}
\end{equation}%
with $r=(i\beta _{2}^{\ast }\lambda _{2}-\beta _{1}^{\ast }\lambda _{1})\cos
(\varphi )+(i\beta _{2}\lambda _{2}-\beta _{1}\lambda _{1})\sin (\varphi )$
and $k=1-\left\vert \beta _{1}\right\vert ^{2}-\left\vert \beta
_{2}\right\vert ^{2}$. The the third term $h_{2}$, which scales as $\mathcal{%
O}(1)$ in terms of $N$, contains only quadratic terms of the bosonic
operators, and thus governs the quantum fluctuations.

In general, a quadratic Hamiltonian $\hat{h}_{2}$ of $n$ bosonic modes can
be expressed as%
\begin{equation}
\hat{h}_{2}=\underline{a}^{\dag }\mathbf{H}\underline{a}+\underline{a}%
\mathbf{K}\underline{a}+\underline{a}^{\dag }\mathbf{K}^{\ast }\underline{a}%
^{\dag },  \label{HamG}
\end{equation}%
where $\underline{a}=(\hat{a}_{1},\hat{a}_{2},..,\hat{a}_{n})^{\text{T}}$ is
the basis of the $n-$dimentional Hilbert space, and the $n\times n$ matrices
$\mathbf{H}$ and $\mathbf{K}$ satisfy $\mathbf{H}^{\dag }=\mathbf{H}$ and $%
\mathbf{K=K}^{\text{T}}$. Under the basis of $\underline{a}=(\hat{c},\hat{d}%
_{1},\hat{d}_{2})^{\text{T}}$ and $\underline{a}=(\hat{c},\hat{d}_{0})^{%
\text{T}}$, the $3\times 3$ matrices $\mathbf{H}_{\text{N/S}}$ and $\mathbf{K%
}_{\text{N/S}}$ for the NP and SP, and the $2\times 2$ matrices $\mathbf{H}_{%
\text{I}}$ and $\mathbf{K}_{\text{I}}$ for the inverted state can be
respectively obtained as,

\begin{widetext}
\begin{equation}
\mathbf{H}_{\text{N/S}}=\left(
\begin{array}{ccc}
\hbar \omega & G_{2,\varphi }+J_{1,\varphi } & iG_{1,\varphi }-iJ_{2,\varphi
} \\
G_{2,\varphi }^{\ast }+J_{1,\varphi }^{\ast } & -D_{1}-B_{1}+\hbar \omega
_{0} & Y_{h} \\
-iG_{1,\varphi }^{\ast }+iJ_{2,\varphi }^{\ast } & Y_{h}^{\ast } &
-iD_{2}+iB_{2}+\hbar \omega _{0}%
\end{array}%
\right) ,
\end{equation}

\begin{equation}
\mathbf{K}_{\text{N/S}}=\left(
\begin{array}{ccc}
0 & -G_{2,\frac{\pi }{2}-\varphi }^{\ast }+J_{1,\frac{\pi }{2}-\varphi
}^{\ast } & -iG_{1,\frac{\pi }{2}-\varphi }^{\ast }-iJ_{2,\frac{\pi }{2}%
-\varphi }^{\ast } \\
-G_{2,\frac{\pi }{2}-\varphi }^{\ast }+J_{1,\frac{\pi }{2}-\varphi }^{\ast }
& S_{1}+X_{1} & Y_{k} \\
-iG_{1,\frac{\pi }{2}-\varphi }^{\ast }-iJ_{2,\frac{\pi }{2}-\varphi }^{\ast
} & Y_{k} & iS_{1}-iX_{2}%
\end{array}%
\right) ,
\end{equation}

\begin{equation}
\mathbf{H}_{\text{I}}=\left(
\begin{array}{cc}
\hbar \omega & -\hbar \cos (\varphi )(i\eta _{2}e^{-i\theta }-\eta _{1}) \\
\hbar \cos (\varphi )(i\eta _{2}e^{i\theta }+\eta _{1}) & -\hbar \omega _{0}%
\end{array}%
\right) ,
\end{equation}

and

\begin{equation}
\mathbf{K}_{\text{I}}=\left(
\begin{array}{cc}
0 & \frac{1}{2}\hbar \sin (\varphi )(-i\eta _{2}e^{i\theta }+\eta _{1}) \\
\frac{1}{2}\hbar \sin (\varphi )(-i\eta _{2}e^{i\theta }+\eta _{1}) & 0%
\end{array}%
\right)
\end{equation}%
\end{widetext}

where%
\begin{widetext}
\begin{eqnarray*}
J_{1,\varphi } &=&(-\beta _{1}^{\ast 2}\cos (\varphi )-\sin (\varphi
)\left\vert \beta _{1}\right\vert ^{2}+2\sin (\varphi )k)\lambda _{1}\hbar
/(2\sqrt{k}), \\
J_{2,\varphi } &=&(-\beta _{2}^{\ast 2}\cos (\varphi )-\sin (\varphi
)\left\vert \beta _{2}\right\vert ^{2}+2\sin (\varphi )k)\lambda _{2}\hbar
/(2\sqrt{k}), \\
G_{1,\varphi } &=&(i\cos (\varphi )\beta _{2}^{\ast }\beta _{1}^{\ast
}+i\sin (\varphi )\beta _{2}^{\ast }\beta _{1})\lambda _{1}\hbar /(2\sqrt{k}%
), \\
G_{2,\varphi } &=&(i\cos (\varphi )\beta _{2}^{\ast }\beta _{1}^{\ast
}+i\sin (\varphi )\beta _{1}^{\ast }\beta _{2})\lambda _{2}\hbar /(2\sqrt{k}%
), \\
D_{1} &=&[(-2i\beta _{2}^{\ast }\alpha ^{\ast }k\lambda _{2}-i\beta
_{2}^{\ast }\left\vert \beta _{1}\right\vert ^{2}\alpha ^{\ast }\lambda
_{2}+\left\vert \beta _{1}\right\vert ^{2}\beta _{1}\alpha \lambda
_{1}+4\alpha \beta _{1}k\lambda _{1})\cos (\varphi )+(-2i\alpha ^{\ast
}\beta _{2}\lambda _{2}k+4\alpha \beta _{1}^{\ast }\lambda _{1}k)\sin
(\varphi )]/(4k^{3/2}), \\
D_{2} &=&[(-2i\beta _{1}^{\ast }\alpha ^{\ast }k\lambda _{1}-i\beta
_{1}^{\ast }\left\vert \beta _{2}\right\vert ^{2}\alpha ^{\ast }\lambda
_{1}+\left\vert \beta _{2}\right\vert ^{2}\beta _{2}\alpha \lambda
_{2}+4\alpha \beta _{2}k\lambda _{2})\cos (\varphi )+(-2i\alpha ^{\ast
}\beta _{1}\lambda _{1}k+4\alpha \beta _{2}^{\ast }\lambda _{2}k)\sin
(\varphi )]/(4k^{3/2}), \\
B_{1} &=&[(2i\beta _{2}^{\ast }\alpha k\lambda _{2}-i\left\vert \beta
_{1}\right\vert ^{2}\beta _{2}\alpha ^{\ast }\lambda _{2}+\left\vert \beta
_{1}\right\vert ^{2}\beta _{1}\alpha \lambda _{1}+i\left\vert \beta
_{1}\right\vert ^{2}\beta _{2}\alpha \lambda _{2}+\left\vert \beta
_{1}\right\vert ^{2}\beta _{1}\alpha ^{\ast }\lambda _{1})\sin (\varphi ) \\
&&+(i\left\vert \beta _{1}\right\vert ^{2}\beta _{2}\alpha \lambda
_{2}+2i\beta _{2}\alpha k\lambda _{2}+\left\vert \beta _{1}\right\vert
^{2}\beta _{1}^{\ast }\alpha ^{\ast }\lambda _{1})\cos (\varphi
)]/(4k^{3/2}), \\
B_{2} &=&[(2i\beta _{1}^{\ast }\alpha k\lambda _{1}-i\left\vert \beta
_{2}\right\vert ^{2}\beta _{1}\alpha ^{\ast }\lambda _{1}+\left\vert \beta
_{2}\right\vert ^{2}\beta _{2}\alpha \lambda _{2}+i\left\vert \beta
_{2}\right\vert ^{2}\beta _{1}\alpha \lambda _{1}+\left\vert \beta
_{2}\right\vert ^{2}\beta _{2}\alpha ^{\ast }\lambda _{2})\sin (\varphi ) \\
&&+(i\left\vert \beta _{2}\right\vert ^{2}\beta _{1}\alpha \lambda
_{1}+2i\beta _{1}\alpha k\lambda _{1}+\left\vert \beta _{2}\right\vert
^{2}\beta _{2}^{\ast }\alpha ^{\ast }\lambda _{2})\cos (\varphi
)]/(4k^{3/2}), \\
S_{1} &=&\hbar \beta _{1}[(i\alpha ^{\ast }\beta _{1}\beta _{2}\lambda
_{2}-\alpha ^{\ast }\beta _{1}^{2}\lambda _{1}-\left\vert \beta
_{1}\right\vert ^{2}\alpha \lambda _{1}-4\alpha k\lambda _{1})\sin (\varphi
)+(i\alpha ^{\ast }\beta _{2}^{\ast }\beta _{1}\lambda _{2}-\alpha \beta
_{1}^{2}\lambda _{1})\cos (\varphi )]/(8k^{3/2}), \\
S_{2} &=&\hbar \beta _{2}[(i\alpha ^{\ast }\beta _{2}\beta _{1}\lambda
_{1}-\alpha ^{\ast }\beta _{2}^{2}\lambda _{2}-\left\vert \beta
_{2}\right\vert ^{2}\alpha \lambda _{2}-4\alpha k\lambda _{2})\sin (\varphi
)+(i\alpha ^{\ast }\beta _{1}^{\ast }\beta _{2}\lambda _{1}-\alpha \beta
_{2}^{2}\lambda _{2})\cos (\varphi )]/(8k^{3/2}), \\
X_{1} &=&-\hbar \beta _{1}[(i\beta _{2}\beta _{1}\alpha \lambda _{2}+\alpha
^{\ast }\left\vert \beta _{1}\right\vert ^{2}\lambda _{1}+4\alpha ^{\ast
}k\lambda _{1})\cos (\varphi )+i\beta _{2}^{\ast }\beta _{1}\alpha \lambda
_{2}]/(8k^{3/2}), \\
X_{2} &=&-\hbar \beta _{2}[(i\beta _{1}\beta _{2}\alpha \lambda _{1}+\alpha
^{\ast }\left\vert \beta _{2}\right\vert ^{2}\lambda _{2}+4\alpha ^{\ast
}k\lambda _{2})\cos (\varphi )+i\beta _{1}^{\ast }\beta _{2}\alpha \lambda
_{1}]/(8k^{3/2}), \\
\eta _{1} &=&\sqrt{N_{1}/N}\lambda _{1},\text{and }\eta _{2}=\lambda _{2}%
\sqrt{1-N_{1}/N}.
\end{eqnarray*}%
\end{widetext}

\section{Eigenstates and the excitation spectra in the closed system}

In the closed system, the solutions of the expectation values $\alpha $ and $%
\beta _{1,2}$ are determined by the extrema of the ME Eq.~(\ref{Me}). We aim
to obtain the expression of the ME in terms of $\alpha $ and $\alpha ^{\ast
} $, from which the energy landscape can be shown clearly. To this end, the
equilibrium condition $\partial E/\partial Z=0$ ($Z=\beta _{1,2}$,$\beta
_{1,2}^{\ast }$) should be applied, yielding four equations%
\begin{equation}
(-\lambda _{1}\cos (\varphi )\alpha ^{\ast }-\lambda _{1}\sin (\varphi
)\alpha )\sqrt{k}-\frac{(r\alpha ^{\ast }+r^{\ast }\alpha )\beta _{1}}{2%
\sqrt{k}}-\beta _{1}\omega _{0}=0,  \label{Eq1}
\end{equation}%
\begin{equation}
(-\lambda _{1}\cos (\varphi )\alpha ^{\ast }-\lambda _{1}\sin (\varphi
)\alpha )\sqrt{k}-\frac{(r\alpha ^{\ast }+r^{\ast }\alpha )\beta _{1}}{2%
\sqrt{k}}-\beta _{1}\omega _{0}=0,  \label{Eq2}
\end{equation}%
\begin{equation}
(i\lambda _{2}\cos (\varphi )\alpha ^{\ast }-i\lambda _{2}\sin (\varphi
)\alpha )\sqrt{k}-\frac{(r\alpha ^{\ast }+r^{\ast }\alpha )\beta _{2}}{2%
\sqrt{k}}-\beta _{2}\omega _{0}=0,  \label{Eq3}
\end{equation}%
\begin{equation}
(i\lambda _{2}\sin (\varphi )\alpha ^{\ast }-i\lambda _{2}\cos (\varphi
)\alpha )\sqrt{k}-\frac{(r\alpha ^{\ast }+r^{\ast }\alpha )\beta _{2}^{\ast }%
}{2\sqrt{k}}-\beta _{2}^{\ast }\omega _{0}=0.  \label{Eq4}
\end{equation}

After some algebraic manipulations on Eqs.~(\ref{Eq1})-(\ref{Eq4}), we have
\begin{equation}
r\alpha ^{\ast }+r^{\ast }\alpha =\frac{2\omega _{0}\sqrt{k}(k-1)}{1-2k}
\label{SE1}
\end{equation}%
and%
\begin{eqnarray}
&&(\cos (\varphi )\sin (\varphi )(\lambda _{1}^{2}-\lambda _{2}^{2})(\alpha
^{2}+\alpha ^{\ast 2})+(\lambda _{1}^{2}+\lambda _{2}^{2})\left\vert \alpha
\right\vert ^{2})k  \notag \\
&=&\left( \frac{r\alpha ^{\ast }+r^{\ast }\alpha }{2\sqrt{k}}+\omega
_{0}\right) (1-k).  \label{SE2}
\end{eqnarray}%
Eliminating the variables $r$ and $k$ in Eq.~(\ref{Me}) by using Eqs.~(\ref%
{SE1})-(\ref{SE2}), the ME can be expressed in terms of $\alpha $ and $%
\alpha ^{\ast }$ as%
\begin{equation}
E=\frac{N\hbar \omega _{0}\sqrt{q+\omega _{0}^{2}}(\omega _{0}-2\omega
\left\vert \alpha \right\vert ^{2})-qN\hbar \omega _{0}}{2\omega _{0}\sqrt{%
q+\omega _{0}^{2}}}  \label{Me2}
\end{equation}%
where $q=4\mathcal{B}(\alpha ^{\ast 2}+\alpha ^{2})+4L\left\vert \alpha
\right\vert ^{2}$, with $\mathcal{B}=\cos (\varphi )\sin (\varphi )(\lambda
_{1}^{2}-\lambda _{2}^{2})$ and $L=\lambda _{1}^{2}+\lambda _{2}^{2}$.

Depending on the values of $\mathcal{B}$ and $L$, the ME is minimized by one
trivial solution [NP in case (i)] and three different nontrivial solutions
[SP in cases (ii-iv)],

(i) $\alpha =0$ for $(2\left\vert \mathcal{B}\right\vert +L)^{2}<\omega
^{2}\omega _{0}^{2}$, (ii) $\alpha =\pm \sqrt{\lbrack (2\mathcal{B}%
+L)^{2}-\omega ^{2}\omega _{0}^{2}]/4(2\mathcal{B}+L)\omega ^{2}}$ for $%
(2\left\vert \mathcal{B}\right\vert +L)^{2}\geqslant \omega ^{2}\omega
_{0}^{2}$ and $\mathcal{B}>0$, (iii) $\alpha =\pm \sqrt{\lbrack (2\mathcal{B}%
-L)^{2}-\omega ^{2}\omega _{0}^{2}]/4(2\mathcal{B}-L)\omega ^{2}}$ for $%
(2\left\vert \mathcal{B}\right\vert +L)^{2}\geqslant \omega ^{2}\omega
_{0}^{2}$ and $\mathcal{B}<0$, and (iv) $\alpha =\left\vert \alpha
\right\vert e^{i\phi }$ with $\left\vert \alpha \right\vert =\sqrt{%
L(L^{2}-\omega ^{2}\omega _{0}^{2})}/(2L\omega )$, for $L^{2}\geqslant
\omega ^{2}\omega _{0}^{2}$ and $\mathcal{B}=0$.

The imaginary (real) part of $\alpha $ is zero for case (ii) [(iii)], and
the sign prefactor of $\alpha $ indicates the $%
\mathbb{Z}
_{2}$ symmetry of the Hamiltonian. Note that case (iv) represents a class of
continuous solutions characterized by the phase $\phi $, which signals the
breaking of the U(1) symmetry. This is consistent with the fact that the ME in
Eq.~(\ref{Me2}) is free of any phase rotation of $\alpha $ for $\mathcal{B}%
=0 $.
\begin{figure}[tp]
\includegraphics[width=8.0cm]{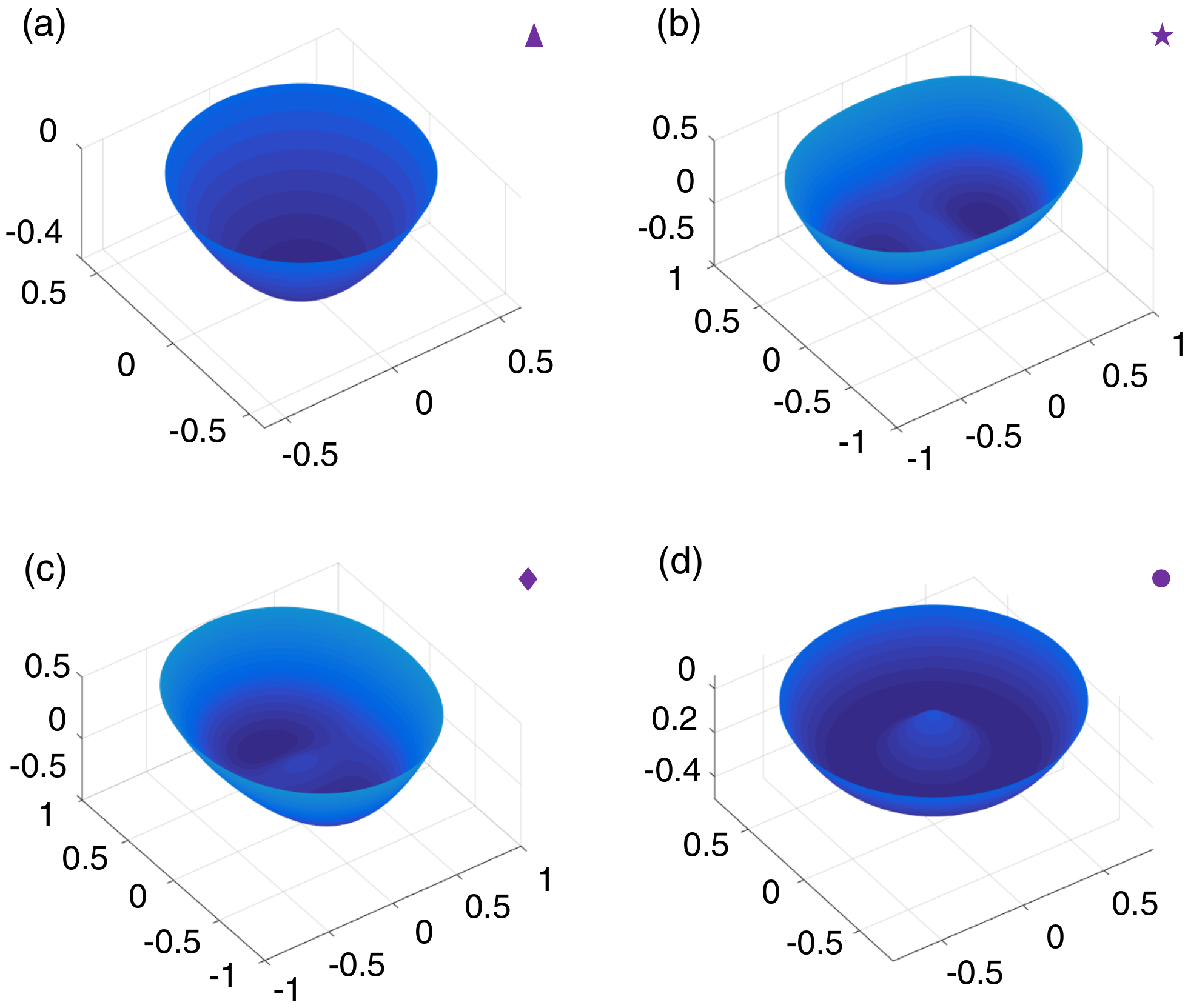}
\caption{The mean-field landscapes for the corresponding parameter locations
indicated by the symbols of (a) triangle, (b) star, (c) diamond, and (d)\
round in Fig. \protect\ref{closed}(c). }
\label{ME}
\end{figure}

With the solutions of $\alpha $, the other two order parameters $\beta _{1}$
and $\beta _{2}$ can be straightforwardly derived by employing Eqs.~(\ref%
{Eq1})-(\ref{SE2}). The complete expressions of $\beta _{1,2}$ are, however,
extremely lengthy and we thus do not list them\ here.

The mean-field solutions are stable only if their associated excitation
energies are real. For systems with $n$ bosonic modes, the excitation
spectra are obtained by diagonalizing the Hopfield-Bogoliubov matrix \cite%
{superradiance4},
\begin{equation}
D_{\text{H}}=\left(
\begin{array}{cc}
\mathbf{H} & \mathbf{K} \\
-\mathbf{K}^{\dag } & -\mathbf{H}^{\text{T}}%
\end{array}%
\right)  \label{HB}
\end{equation}%
where $\mathbf{H}$ and $\mathbf{K}$\ are the $n\times n$ matrix\ defined in
the Appendix B. For the NP and SP considered in the present system, the
diagonalization of the Hopfield-Bogoliubov matrix~(\ref{HB}) produces $6$
eigenfrequencies, which are paired with opposite signs $\pm \omega _{i}$ ($%
i=1,2,3$). The solutions of $\alpha $ and $\beta _{1,2}$, together with
their associated eigenfrequencies, determine the whole closed phase diagram.

Figure~\ref{ME} plot the ME landscapes for four representative points in the
$\lambda _{1}-\lambda _{2}$ parameter space indicated by the symbols of
triangle, star, diamond and round in Fig.~\ref{closed}(c). It should be
noticed that, unlike the NP and SP which minimize $E$, the e-NP [Fig.~\ref%
{ME}(b)] corresponds to a local maximum of the ME.

\section{Steady states and the stability analysis in the open system}

In this Section, we detail the derivation of the steady-state solutions of
the master equation $\partial _{t}\hat{\rho}=\mathcal{\hat{L}}\hat{\rho}=0$,
with which the HP Hamiltonian in the open system is obtained.\ We remark
that in the open system, the HP Hamiltonians for the NP and inverted state
are the same as those of the closed system, whereas they have a different
form for the SP.

\subsection{Superradiant steady state}

Utilizing a mean-field decoupling by equating the cavity field operator $%
\hat{a}$ with its expectation value $\left\langle \hat{a}\right\rangle $,
the Hamiltonian~(\ref{Ham}) can be written as%
\begin{eqnarray}
H &=&\hbar \omega _{0}(\hat{\Lambda}_{1,1}+\hat{\Lambda}_{2,2})+M_{1}\hat{%
\Lambda}_{1,0}+M_{2}\hat{\Lambda}_{2,0}  \notag \\
&&+M_{1}^{\ast }\hat{\Lambda}_{0,1}+M_{2}^{\ast }\hat{\Lambda}_{0,2}
\label{MH}
\end{eqnarray}%
where

\begin{widetext}
\begin{eqnarray}
M_{1} &=&\frac{\hbar \lambda _{1}}{N(\kappa ^{2}+\omega ^{2})}\{[\omega
\left\langle \hat{\Lambda}_{0,1}\right\rangle +i\kappa \left\langle \hat{%
\Lambda}_{0,1}\right\rangle +2\sin (\varphi )\cos (\varphi )\omega
\left\langle \hat{\Lambda}_{1,0}\right\rangle -2i\cos ^{2}(\varphi )\kappa
\left\langle \hat{\Lambda}_{0,1}\right\rangle ]\lambda _{1}  \notag \\
&&+[\kappa \left\langle \hat{\Lambda}_{0,2}\right\rangle -i\omega
\left\langle \hat{\Lambda}_{0,2}\right\rangle +2\sin (\varphi )\cos (\varphi
)\kappa \left\langle \hat{\Lambda}_{2,0}\right\rangle +2i\cos ^{2}(\varphi
)\omega \left\langle \hat{\Lambda}_{0,2}\right\rangle ]\lambda _{2}\},
\label{M1}
\end{eqnarray}

\begin{eqnarray}
M_{2} &=&\frac{i\hbar \lambda _{2}}{N(\kappa ^{2}+\omega ^{2})}\{[-2\omega
\left\langle \hat{\Lambda}_{0,1}\right\rangle \cos ^{2}(\varphi )+2i\sin
(\varphi )\cos (\varphi )\kappa \left\langle \hat{\Lambda}%
_{1,0}\right\rangle +(\omega +i\kappa )\left\langle \hat{\Lambda}%
_{0,1}\right\rangle ]\lambda _{1}  \notag \\
&&+[-2\kappa \left\langle \hat{\Lambda}_{0,2}\right\rangle \cos ^{2}(\varphi
)-2i\omega \sin (\varphi )\cos (\varphi )\left\langle \hat{\Lambda}%
_{2,0}\right\rangle -(i\omega -\kappa )\left\langle \hat{\Lambda}%
_{0,2}\right\rangle ]\lambda _{2}\}.  \label{M2}
\end{eqnarray}%
\end{widetext}

Note that in writing Eq.~(\ref{MH}), the steady state of the cavity field
\begin{eqnarray}
\left\langle \hat{a}\right\rangle &=&\frac{1}{(\omega +i\kappa )\sqrt{N}}%
\{[\cos (\varphi )\left\langle \hat{\Lambda}_{1,0}\right\rangle +\sin
(\varphi )\left\langle \hat{\Lambda}_{0,1}\right\rangle ]\lambda _{1}  \notag
\\
&&-[i\cos (\varphi )\left\langle \hat{\Lambda}_{2,0}\right\rangle +i\sin
(\varphi )\left\langle \hat{\Lambda}_{0,2}\right\rangle ]\lambda _{2}\}
\end{eqnarray}%
has been used. The Hamiltonian~(\ref{MH}) produces the equations of motion
for the atomic operators $\hat{\Lambda}_{i,j}$ ($i,j=1,2,3$), which are
solved under the constraint of the SU(3) atomic symmetry \cite%
{ThreeDicke4,ThreeDicke8}, i.e.,%
\begin{equation}
\sum_{\mu =0}^{2}\left\langle \hat{\Lambda}_{\mu ,\mu }\right\rangle =N,
\end{equation}

\begin{widetext}
\begin{equation}
\sum_{\mu =0}^{2}\left\langle \hat{\Lambda}_{\mu ,\mu }\right\rangle
^{2}+\sum_{\{\mu ,\nu \}}\left( 3\left\vert \left\langle \hat{\Lambda}_{\mu
,\nu }\right\rangle \right\vert ^{2}-\left\langle \hat{\Lambda}_{\mu ,\mu
}\right\rangle \left\langle \hat{\Lambda}_{\nu ,\nu }\right\rangle \right)
=N^{2},
\end{equation}%
and

\begin{equation}
\frac{9}{2}\sum_{\{\mu ,\nu ,\rho \}}\left\vert \left\langle \hat{\Lambda}%
_{\mu ,\nu }\right\rangle \right\vert ^{2}\left( \left\langle \hat{\Lambda}%
_{\mu ,\mu }\right\rangle +\left\langle \hat{\Lambda}_{\nu ,\nu
}\right\rangle -2\left\langle \hat{\Lambda}_{\rho ,\rho }\right\rangle
\right) -\frac{1}{2}\prod_{\{\mu ,\nu ,\rho \}}\left( \left\langle \hat{%
\Lambda}_{\mu ,\mu }\right\rangle +\left\langle \hat{\Lambda}_{\nu ,\nu
}\right\rangle -2\left\langle \hat{\Lambda}_{\rho ,\rho }\right\rangle
\right) +27\left\vert \left\langle \hat{\Lambda}_{0,1}\right\rangle
\left\langle \hat{\Lambda}_{1,2}\right\rangle \left\langle \hat{\Lambda}%
_{2,0}\right\rangle \right\vert =N^{3},
\end{equation}%
\end{widetext}

where the summation $\sum_{\{\mu ,\nu \}}$ runs over the pairs $\{\mu ,\nu
\}=\{0,1\},\{1,2\},\{2,0\}$, while the summation $\sum_{\{\mu ,\nu ,\rho \}}$
and the product $\prod_{\{\mu ,\nu ,\rho \}}$ run over the triplets $\{\mu
,\nu ,\rho \}=\{0,1,2\},\{1,2,0\},\{2,0,1\}$. The solutions reads

\begin{widetext}
\begin{eqnarray}
\left\langle \hat{\Lambda}_{0,1}\right\rangle  &=&\frac{NM_{1}}{\sqrt{%
4\left\vert M_{1}\right\vert ^{2}+4\left\vert M_{2}\right\vert ^{2}+\hbar
^{2}\omega _{0}^{2}}},  \label{G01} \\
\left\langle \hat{\Lambda}_{0,2}\right\rangle  &=&\frac{NM_{2}}{\sqrt{%
4\left\vert M_{1}\right\vert ^{2}+4\left\vert M_{2}\right\vert ^{2}+\hbar
^{2}\omega _{0}^{2}}},  \label{G02} \\
\left\langle \hat{\Lambda}_{0,0}\right\rangle  &=&\frac{N}{2}-\frac{N\hbar
\omega _{0}}{2\sqrt{4\left\vert M_{1}\right\vert ^{2}+4\left\vert
M_{2}\right\vert ^{2}+\hbar ^{2}\omega _{0}^{2}}}, \\
\left\langle \hat{\Lambda}_{1,1}\right\rangle  &=&\frac{N\left\vert
M_{1}\right\vert ^{2}}{2(\left\vert M_{1}\right\vert ^{2}+\left\vert
M_{2}\right\vert ^{2})}\left( \frac{\hbar \omega _{0}}{\sqrt{4\left\vert
M_{1}\right\vert ^{2}+4\left\vert M_{2}\right\vert ^{2}+\hbar ^{2}\omega
_{0}^{2}}}+1\right) , \\
\left\langle \hat{\Lambda}_{2,2}\right\rangle  &=&\frac{N\left\vert
M_{2}\right\vert ^{2}}{2(\left\vert M_{1}\right\vert ^{2}+\left\vert
M_{2}\right\vert ^{2})}\left( \frac{\hbar \omega _{0}}{\sqrt{4\left\vert
M_{1}\right\vert ^{2}+4\left\vert M_{2}\right\vert ^{2}+\hbar ^{2}\omega
_{0}^{2}}}+1\right) , \\
\left\langle \hat{\Lambda}_{1,2}\right\rangle  &=&\frac{NM_{1}^{\ast }M_{2}}{%
2(\left\vert M_{1}\right\vert ^{2}+\left\vert M_{2}\right\vert ^{2})}\left(
\frac{\hbar \omega _{0}}{\sqrt{4\left\vert M_{1}\right\vert ^{2}+4\left\vert
M_{2}\right\vert ^{2}+\hbar ^{2}\omega _{0}^{2}}}+1\right) ,
\end{eqnarray}%
\end{widetext}

and $\left\langle \hat{\Lambda}_{1,0}\right\rangle =\left\langle \hat{\Lambda%
}_{0,1}\right\rangle ^{\ast }$, $\left\langle \hat{\Lambda}%
_{2,0}\right\rangle =\left\langle \hat{\Lambda}_{0,2}\right\rangle ^{\ast }$
and $\left\langle \hat{\Lambda}_{2,1}\right\rangle =\left\langle \hat{\Lambda%
}_{1,2}\right\rangle ^{\ast }$. Eliminating the variables $\left\langle \hat{%
\Lambda}_{0,1}\right\rangle $, $\left\langle \hat{\Lambda}%
_{0,2}\right\rangle $, $\left\langle \hat{\Lambda}_{1,0}\right\rangle $ and $%
\left\langle \hat{\Lambda}_{2,0}\right\rangle $ in Eqs.~(\ref{M1})-(\ref{M2}%
) by making use of Eqs.~(\ref{G01}) -(\ref{G02}), we have%
\begin{widetext}
\begin{equation}
\frac{M_{1}}{\hbar \lambda _{1}}=\frac{-M_{1}\lambda _{1}(2i\cos
^{2}(\varphi )\kappa -i\kappa -\omega )+M_{2}\lambda _{2}[2i\cos
^{2}(\varphi )\omega -i\omega +\kappa ]+2M_{1}^{\ast }\lambda _{1}\cos
(\varphi )\sin (\varphi )\omega +2M_{2}^{\ast }\lambda _{2}\cos (\varphi
)\sin (\varphi )\kappa }{\sqrt{4\left\vert M_{1}\right\vert ^{2}+4\left\vert
M_{2}\right\vert ^{2}+\hbar ^{2}\omega _{0}^{2}}(\kappa ^{2}+\omega ^{2})}
\label{MS1}
\end{equation}

\begin{equation}
\frac{M_{2}}{\hbar \lambda _{2}}=\frac{-M_{2}\lambda _{2}(2i\cos
^{2}(\varphi )\kappa -i\kappa -\omega )-M_{1}\lambda _{1}[2i\cos
^{2}(\varphi )\omega -i\omega +\kappa ]-2M_{1}^{\ast }\lambda _{1}\cos
(\varphi )\sin (\varphi )\kappa +2M_{2}^{\ast }\lambda _{2}\cos (\varphi
)\sin (\varphi )\omega }{\sqrt{4\left\vert M_{1}\right\vert ^{2}+4\left\vert
M_{2}\right\vert ^{2}+\hbar ^{2}\omega _{0}^{2}}(\kappa ^{2}+\omega ^{2})}
\label{MS2}
\end{equation}
\end{widetext}

By solving Eqs.~(\ref{MS1})-(\ref{MS2}) and their complex conjugated
versions, $M_{1,2}$ can be normally determined. While the expressions of $%
M_{1,2}$ are too lengthy to be listed here, they are related to the order
parameters $\beta _{1,2}$ through the simple algebraic relations%
\begin{eqnarray}
\beta _{1} &=&\frac{\sqrt{2}M_{1}}{\sqrt{\hbar \omega _{0}+\sqrt{4\left\vert
M_{1}\right\vert ^{2}+4\left\vert M_{2}\right\vert ^{2}+\hbar ^{2}\omega
_{0}^{2}}}} \\
\beta _{2} &=&\frac{\sqrt{2}M_{2}}{\sqrt{\hbar \omega _{0}+\sqrt{4\left\vert
M_{1}\right\vert ^{2}+4\left\vert M_{2}\right\vert ^{2}+\hbar ^{2}\omega
_{0}^{2}}}}.
\end{eqnarray}%
With the obtained $\beta _{1,2}$ and taking into consideration the
fluctuation Hamiltonian~(\ref{HamG}), the matrices $\mathbf{H}$ and $\mathbf{%
K}$ are uniquely fixed.

\subsection{Third quantization and the stability analysis}

The third quantization approach exactly solves the Lindblad master equation
for an arbitrary quadratic system of $n$ bosons/fermions with linear bath
operators \cite{ThirdQuant1,ThirdQuant2}, and is hence suitable for the
stability analysis around the obtained non-equilibrium steady states. We
here skip the details of this method in quantizing the density operator, and
focus on the most relevant steps in analyzing the system stability.
\begin{figure}[tp]
\includegraphics[width=8cm]{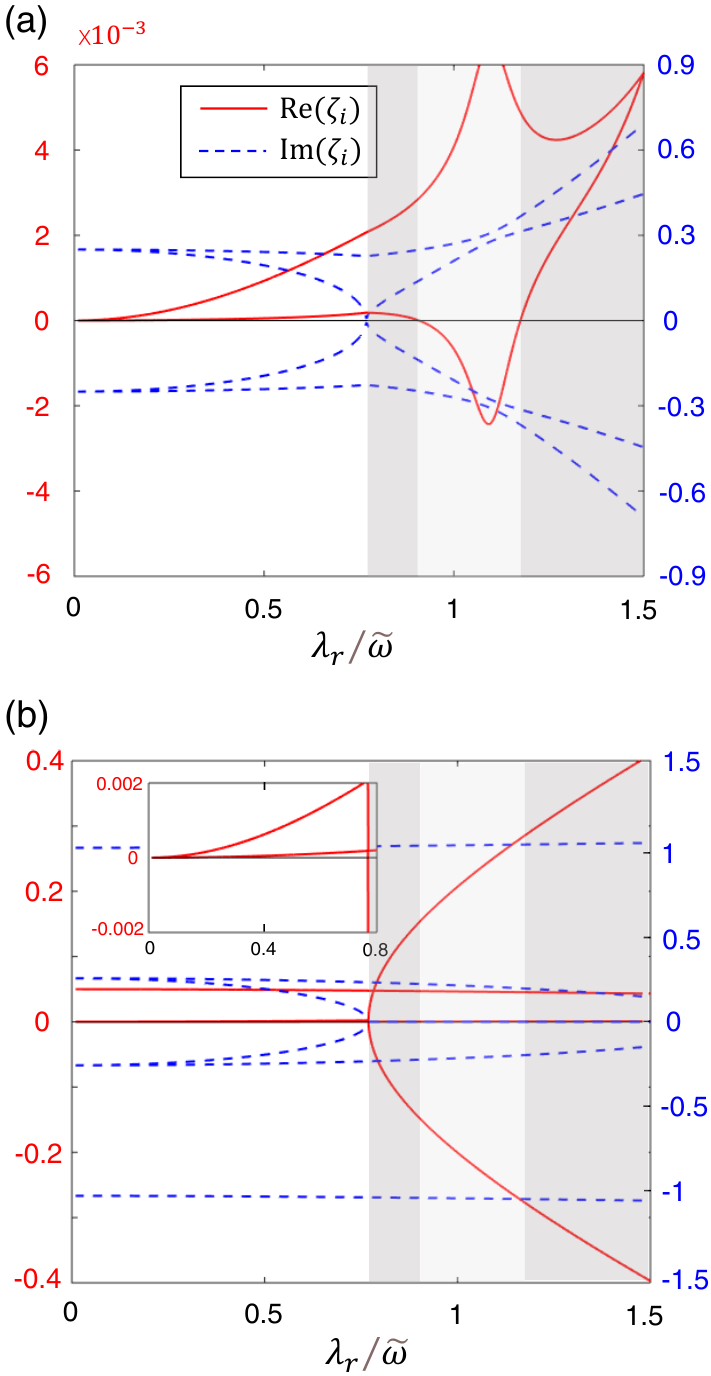}
\caption{The real (red solid) and imaginary (blue dashed) parts of the
rapidities $\protect\zeta _{i}$ on top of the (a) superradiant and (b)
normal phases. As the coupling strength $\protect\lambda _{r}$ increases,
the system traverses the regions of multi-phase coexistence of NP+SP
(white), SP (dark grey), OS (light grey), and SP (dark grey). The parameters are chosen as $%
\protect\varphi =0.22\protect\pi $, $\protect\lambda _{2}/\protect\lambda %
_{1}=0.41$, $\protect\omega =4\protect\omega _{0}=2\tilde{\protect\omega}$
and $\protect\kappa =0.1\tilde{\protect\omega}$.\ }
\label{Rap}
\end{figure}

Under the framework of the third quantization, the dynamical property of the
steady states is captured by the shape matrix of the Liouvillian,%
\begin{equation}
\chi =\frac{1}{2}\left(
\begin{array}{cc}
i\mathbf{H}^{\ast }-\mathbf{N}^{\ast }+\mathbf{M} & -2\mathbf{K}-\mathbf{L}+%
\mathbf{L}^{\text{T}} \\
2i\mathbf{K}^{\ast }-\mathbf{L}^{\ast }+\mathbf{L}^{\dag } & -i\mathbf{H}-%
\mathbf{N}+\mathbf{M}^{\ast }%
\end{array}%
\right)
\end{equation}%
where $\mathbf{H}$ and $\mathbf{K}$\ are defined in the Appendix B, and the
other three matrices are given by%
\begin{equation}
\mathbf{M=}\underline{l}_{1}\otimes \underline{l}_{1}^{\ast }\text{, }%
\mathbf{N=}\underline{l}_{2}\otimes \underline{l}_{2}^{\ast }\text{, }%
\mathbf{L}=\underline{l}_{1}\otimes \underline{l}_{2}^{\ast }.  \label{MNL}
\end{equation}%
The matrices $\underline{l}_{1,2}$ in Eq.~(\ref{MNL}) are defined through
the linear Lindblad bath operators in the form of%
\begin{equation}
L=\underline{l}_{1}\cdot \underline{a}+\underline{l}_{2}\cdot \underline{%
a^{\dag }}.
\end{equation}%
Given that the bath operator for our model is $L=\sqrt{\kappa }\hat{c}$, we
have the operator basis $\underline{a}=(\hat{c},\hat{d}_{1},\hat{d}_{2})^{%
\text{T}}$ and the corresponding matrices $\underline{l}_{1}=(\sqrt{\kappa }%
,0,0)^{\text{T}}$ and $\underline{l}_{2}=(0,0,0)^{\text{T}}$ for the SP and
NP, leading to%
\begin{equation}
\mathbf{M}_{\text{N/S}}=\text{diag}(\kappa ,0,0)\text{, }\mathbf{N}_{\text{%
N/S}}=\mathbf{L}_{\text{N/S}}=\mathbf{0}_{3\times 3}
\end{equation}%
whereas for the inverted state, we have $\underline{a}=(\hat{c},\hat{d}%
_{0})^{\text{T}}$, $\underline{l}_{1}=(\sqrt{\kappa },0)^{\text{T}}$, and$\
\underline{l}_{2}=(0,0)^{\text{T}}$, resulting in
\begin{equation}
\mathbf{M}_{\text{I}}=\text{diag}(\kappa ,0)\text{, }\mathbf{N}_{\text{I}}=%
\mathbf{L}_{\text{I}}=\mathbf{0}_{2\times 2}.
\end{equation}

The eigenvalues of the shape matrix $\chi $, dubbed rapidities and
represented by $\zeta _{i}$, are negatively related to the eigenvalues of
the Liouvillian and thus play the role of excitation energies in the closed
system. It follows that the real part of $\zeta _{i}$ determines the
stability of the corresponding steady state and, the imaginary part
represents the oscillation frequency of the fluctuations. The steady state
is stable if and only if the real part of all the rapidities are
nonnegative, i.e., min(Re$\zeta _{i}$) $\geqslant 0$. For parameters region
where both NP and SP are unstable, we should further integrate the equations
of motion starting from arbitrary initial conditions to identify possible
limit-cycle attractors. In Fig.~\ref{Rap}, we plot the rapidities on top of
the NP and SP for some representative parameters.

\section{Mean-field equations of motion}

According to the master equation $\partial _{t}\hat{\rho}=\mathcal{\hat{L}}%
\hat{\rho}$, we can obtain the equation of motion for the expectation of a
general operator $\mathcal{\hat{O}}$,%
\begin{equation}
\frac{d}{dt}\left\langle \mathcal{\hat{O}}\right\rangle =-\frac{i}{\hbar }%
\left\langle \left[ \mathcal{\hat{O}},\hat{H}\right] \right\rangle -\kappa
\left\{ \left\langle \left[ \mathcal{\hat{O}},\hat{a}^{\dag }\right] \hat{a}%
\right\rangle -\left\langle \hat{a}^{\dag }\left[ \mathcal{\hat{O}},\hat{a}%
\right] \right\rangle \right\} .
\end{equation}%
For our model, the operator $\mathcal{\hat{O}}$ is chosen as the pseudospin
operators $\hat{\Lambda}_{i,j}$ and cavity field operator $\hat{a}$.
Applying the mean-field decoupling $\left\langle \hat{\Lambda}_{i,j}\hat{a}%
\right\rangle \approx \left\langle \hat{\Lambda}_{i,j}\right\rangle
\left\langle \hat{a}\right\rangle $, we can derive the closed set of
equations of motion

\begin{widetext}
\begin{eqnarray*}
\frac{d}{dt}\left\langle \hat{a}\right\rangle  &=&(i\hbar \omega -\kappa
)\left\langle \hat{a}\right\rangle -\frac{i\hbar \lambda _{1}[\cos (\varphi
)\left\langle \hat{\Lambda}_{1,0}\right\rangle +\sin (\varphi )\left\langle
\hat{\Lambda}_{0,1}\right\rangle ]-i\hbar \lambda _{2}[\cos (\varphi
)\left\langle \hat{\Lambda}_{2,0}\right\rangle +\sin (\varphi )\left\langle
\hat{\Lambda}_{0,2}\right\rangle ]}{\sqrt{N}} \\
\frac{d}{dt}\left\langle \hat{a}^{\dag }\right\rangle  &=&(-i\hbar \omega
-\kappa )\left\langle \hat{a}\right\rangle +\frac{i\hbar \lambda _{1}[\cos
(\varphi )\left\langle \hat{\Lambda}_{0,1}\right\rangle +\sin (\varphi
)\left\langle \hat{\Lambda}_{1,0}\right\rangle ]-i\hbar \lambda _{2}[\cos
(\varphi )\left\langle \hat{\Lambda}_{0,2}\right\rangle +\sin (\varphi
)\left\langle \hat{\Lambda}_{2,0}\right\rangle ]}{\sqrt{N}} \\
\frac{d}{dt}\left\langle \hat{\Lambda}_{0,0}\right\rangle  &=&\frac{-i\hbar
\lambda _{1}\{-\left\langle \hat{\Lambda}_{0,1}\right\rangle [\sin (\varphi
)\left\langle \hat{a}\right\rangle +\cos (\varphi )\left\langle \hat{a}%
^{\dag }\right\rangle ]+\left\langle \hat{\Lambda}_{1,0}\right\rangle [\sin
(\varphi )\left\langle \hat{a}^{\dag }\right\rangle +\cos (\varphi
)\left\langle \hat{a}\right\rangle ]\}}{\sqrt{N}} \\
&&+\frac{\hbar \lambda _{2}\{-\left\langle \hat{\Lambda}_{2,0}\right\rangle
[\sin (\varphi )\left\langle \hat{a}\right\rangle -\cos (\varphi
)\left\langle \hat{a}^{\dag }\right\rangle ]+\left\langle \hat{\Lambda}%
_{0,2}\right\rangle [-\sin (\varphi )\left\langle \hat{a}^{\dag
}\right\rangle +\cos (\varphi )\left\langle \hat{a}\right\rangle ]\}}{\sqrt{N%
}} \\
\frac{d}{dt}\left\langle \hat{\Lambda}_{1,1}\right\rangle  &=&\frac{-i\hbar
\lambda _{1}\{\left\langle \hat{\Lambda}_{1,0}\right\rangle [\sin (\varphi
)\left\langle \hat{a}\right\rangle +\cos (\varphi )\left\langle \hat{a}%
^{\dag }\right\rangle ]-\left\langle \hat{\Lambda}_{0,1}\right\rangle [\sin
(\varphi )\left\langle \hat{a}^{\dag }\right\rangle +\cos (\varphi
)\left\langle \hat{a}\right\rangle ]\}}{\sqrt{N}} \\
\frac{d}{dt}\left\langle \hat{\Lambda}_{2,2}\right\rangle  &=&\frac{\hbar
\lambda _{2}\{\left\langle \hat{\Lambda}_{2,0}\right\rangle [\sin (\varphi
)\left\langle \hat{a}\right\rangle -\cos (\varphi )\left\langle \hat{a}%
^{\dag }\right\rangle ]-\left\langle \hat{\Lambda}_{0,2}\right\rangle [-\sin
(\varphi )\left\langle \hat{a}^{\dag }\right\rangle +\cos (\varphi
)\left\langle \hat{a}\right\rangle ]\}}{\sqrt{N}} \\
\frac{d}{dt}\left\langle \hat{\Lambda}_{1,2}\right\rangle  &=&\frac{i\hbar
\lambda _{1}\left\langle \hat{\Lambda}_{0,2}\right\rangle [\sin (\varphi
)\left\langle \hat{a}^{\dag }\right\rangle +\cos (\varphi )\left\langle \hat{%
a}\right\rangle ]+\hbar \lambda _{2}\left\langle \hat{\Lambda}%
_{1,0}\right\rangle [\sin (\varphi )\left\langle \hat{a}\right\rangle -\cos
(\varphi )\left\langle \hat{a}^{\dag }\right\rangle ]}{\sqrt{N}} \\
\frac{d}{dt}\left\langle \hat{\Lambda}_{0,1}\right\rangle  &=&-i\hbar \omega
_{0}\left\langle \hat{\Lambda}_{0,1}\right\rangle -\frac{i\hbar \lambda
_{1}(\left\langle \hat{\Lambda}_{0,0}\right\rangle -\left\langle \hat{\Lambda%
}_{1,1}\right\rangle )[\sin (\varphi )\left\langle \hat{a}\right\rangle
+\cos (\varphi )\left\langle \hat{a}^{\dag }\right\rangle ]}{\sqrt{N}}-\frac{%
\hbar \lambda _{2}\left\langle \hat{\Lambda}_{2,1}\right\rangle [\sin
(\varphi )\left\langle \hat{a}\right\rangle -\cos (\varphi )\left\langle
\hat{a}^{\dag }\right\rangle ]}{\sqrt{N}} \\
\frac{d}{dt}\left\langle \hat{\Lambda}_{0,2}\right\rangle  &=&-i\hbar \omega
_{0}\left\langle \hat{\Lambda}_{0,2}\right\rangle +\frac{\hbar \lambda
_{2}(\left\langle \hat{\Lambda}_{0,0}\right\rangle -\left\langle \hat{\Lambda%
}_{2,2}\right\rangle )[\sin (\varphi )\left\langle \hat{a}\right\rangle
-\cos (\varphi )\left\langle \hat{a}^{\dag }\right\rangle ]}{\sqrt{N}}+\frac{%
i\hbar \lambda _{1}\left\langle \hat{\Lambda}_{1,2}\right\rangle [\sin
(\varphi )\left\langle \hat{a}\right\rangle +\cos (\varphi )\left\langle
\hat{a}^{\dag }\right\rangle ]}{\sqrt{N}} \\
\frac{d}{dt}\left\langle \hat{\Lambda}_{2,1}\right\rangle  &=&\frac{-i\hbar
\lambda _{1}\left\langle \hat{\Lambda}_{2,0}\right\rangle [\sin (\varphi
)\left\langle \hat{a}\right\rangle +\cos (\varphi )\left\langle \hat{a}%
^{\dag }\right\rangle ]+\hbar \lambda _{2}\left\langle \hat{\Lambda}%
_{0,1}\right\rangle [\sin (\varphi )\left\langle \hat{a}^{\dag
}\right\rangle -\cos (\varphi )\left\langle \hat{a}\right\rangle ]}{\sqrt{N}}
\\
\frac{d}{dt}\left\langle \hat{\Lambda}_{1,0}\right\rangle  &=&i\hbar \omega
_{0}\left\langle \hat{\Lambda}_{1,0}\right\rangle +\frac{i\hbar \lambda
_{1}(\left\langle \hat{\Lambda}_{0,0}\right\rangle -\left\langle \hat{\Lambda%
}_{1,1}\right\rangle )[\sin (\varphi )\left\langle \hat{a}^{\dag
}\right\rangle +\cos (\varphi )\left\langle \hat{a}\right\rangle ]}{\sqrt{N}}%
-\frac{\hbar \lambda _{2}\left\langle \hat{\Lambda}_{1,2}\right\rangle [\sin
(\varphi )\left\langle \hat{a}^{\dag }\right\rangle -\cos (\varphi
)\left\langle \hat{a}\right\rangle ]}{\sqrt{N}} \\
\frac{d}{dt}\left\langle \hat{\Lambda}_{2,0}\right\rangle  &=&-i\hbar \omega
_{0}\left\langle \hat{\Lambda}_{2,0}\right\rangle +\frac{\hbar \lambda
_{2}(\left\langle \hat{\Lambda}_{0,0}\right\rangle -\left\langle \hat{\Lambda%
}_{2,2}\right\rangle )[\sin (\varphi )\left\langle \hat{a}^{\dag
}\right\rangle -\cos (\varphi )\left\langle \hat{a}\right\rangle ]}{\sqrt{N}}%
+\frac{i\hbar \lambda _{1}\hat{\Lambda}_{2,1}[\sin (\varphi )\left\langle
\hat{a}^{\dag }\right\rangle +\cos (\varphi )\left\langle \hat{a}%
\right\rangle ]}{\sqrt{N}}
\end{eqnarray*}%
\end{widetext}


\begin{thebibliography}{99}
\bibitem{remark} To be clarity, we further confine the value range of $%
\varphi $ to be $\varphi \in \lbrack 0,\pi /2]$ for the open model. The
parameter $\varphi $ out of this range can be reset by utilizing the
following symmetry transformations: $\varphi \mapsto \varphi +\pi /2$, $%
a\mapsto -a^{\dag }$, $\left\vert 0\right\rangle \mapsto -i\left\vert
0\right\rangle $ and $\lambda _{1(2)}\mapsto \lambda _{2(1)}$.
\end{thebibliography}
\end{document}